\documentclass[11pt]{article}
\pdfoutput=1
\usepackage{appendix}
\usepackage{tikz}
\usetikzlibrary{matrix,arrows,decorations.pathmorphing,decorations.pathreplacing}
\usepackage{jheppub}
\usepackage{amsmath,amssymb,euscript,array,mathrsfs,appendix,ctable,marvosym}
\usepackage{arydshln}
\usepackage{todonotes}
\usepackage{graphicx}
\usepackage[normalem]{ulem}

\input{tcilatex}

\title{Symmetric space $\lambda$-model exchange algebra from 4d holomorphic Chern-Simons theory}
\author{David M. Schmidtt\footnote{david@df.ufscar.br}} 

\affiliation{Departamento de F\'\i sica, Universidade Federal de S\~ao Carlos, \\
Caixa Postal 676, CEP 13565-905, S\~ao Carlos-SP, Brasil} 

\abstract{We derive, within the Hamiltonian formalism, the classical exchange algebra of a lambda deformed string sigma model in a symmetric space directly from a 4d holomorphic Chern-Simons theory. The explicit forms of the extended Lax connection and R-matrix entering the Maillet bracket of the lambda model are explained from a symmetry principle.
This approach, based on a gauge theory, may provide a mechanism for taming the non-ultralocality that afflicts most of the integrable string theories propagating in coset spaces.
\begin{flushleft}
\end{flushleft}
}

\numberwithin{equation}{section}

\begin{document}

\maketitle


\section{Introduction}

In a series of papers \cite{C1,C2,W,CWY1,CWY2,CY}, it was gradually shown that various integrable lattice models and integrable field theories
can be understood as originating from a four-dimensional variant of a Chern-Simons (CS)
theory on the product space $\Sigma\times C$, of a real two-dimensional manifold $\Sigma$ and a
Riemann surface $C$ equipped with a non-vanishing meromorphic 1-form $\omega$. 

Inspired by the framework of \cite{CY}, it was shown in \cite{Vicedo-Holo}, that there is an intimate relation between the 4d CS theory description of \cite{CY} and the description of integrable field theories based on Gaudin models proposed in \cite{V3}. In \cite{Vicedo-Holo}, a 4d CS theory defined on $\Sigma\times \mathbb{CP}^{1}$ was considered and it was shown, within the Hamiltonian formalism that after gauge fixing, the resulting Dirac bracket of the reduced theory takes the Maillet algebra form \cite{Maillet}    
\begin{equation}
\begin{aligned}
\left\{ A_{\sigma }(\sigma ,z)_{\mathbf{1}},A_{\sigma }(\sigma ^{\prime
},z^{\prime })_{\mathbf{2}}\right\} ^{\star} =\big[ R_{%
\mathbf{12}}(z,z^{\prime }),\> &A_{\sigma }(\sigma ,z)_{\mathbf{1}}\big]
\delta _{\sigma \sigma ^{\prime }}-\big[ R_{\mathbf{21}}(z^{\prime },z),A_{\sigma
}(\sigma ^{\prime },z^{\prime })_{\mathbf{2}}\big] \delta _{\sigma
\sigma ^{\prime }} \\
&-\left(R_{\mathbf{12}}(z,z^{\prime })+R_{\mathbf{21}}(z^{\prime },z)\right)\delta _{\sigma \sigma^{\prime}}^{\prime },
\end{aligned} \label{1}
\end{equation}
where
\begin{equation}
R_{\mathbf{12}}(z,z')\sim -\frac{C_{\mathbf{12}}}{z-z'}\varphi(z')^{-1}
\end{equation}
is an R-matrix and $\varphi(z)$ a twist function. The very structure of the R-matrix above suggest that the reduced theory corresponds to an integrable field theory of the principal chiral model type, i.e. a string sigma model propagating in a group manifold $F$ or some sort of deformation thereof determined by the form chosen for $\varphi(z)$. However, the description of a string sigma model in a (semi)-symmetric space $\hat{F}/G$ or a deformation of it, like the one provided by the $\eta$ \cite{Klimcik,eta-def bos,eta-def fer} or $\lambda$ \cite{Sfetsos, lambda-bos,lambda-fer} deformations, was not clear at the time by following the same lines. This was unfortunate, because important non-ultralocal integrable field theories like the $AdS_{5}\times S^{5}$ (super)-string fit precisely in this category. 

It is the purpose of this note to start with a particular 4d CS theory on $\Sigma\times \mathbb{CP}^{1}$ and recover the algebra \eqref{1} for the lambda deformation of a string sigma model propagating in a symmetric space $F/G$, where the R-matrix takes the specific form
\begin{equation}
R_{\mathbf{12}}(z,z^{\prime })=-\frac{2z^{\prime 4}}{z^{4}-z^{\prime 4}}\left( C_{\mathbf{12}}^{(00)}+\frac{z^{2}}{z^{\prime 2}}C_{%
\mathbf{12}}^{(22)}\right)\phi
(z^{\prime })^{-1}. \label{R}
\end{equation}%
Although we have chosen the lambda deformation as a working example, the results can be adapted to include other models with coset structure by considering a different choice of twist function $\phi(z)$ and/or Lie algebra structure. The key strategy employed in order to derive \eqref{R} from first principles, relies in the introduction of an equivariance condition for the CS gauge field that is constructed out of the action of two $\mathbb{Z}_{4}$ cyclic groups. The first one related to the usual $\mathbb{Z}_{4}$ gradation of the Lie algebra $\mathfrak{f}$ that gives the symmetric space structure to $\mathfrak{f}$ and the second one related to the action of the $\mathbb{Z}_{4}$ cyclic group generated by $\rho=i$, acting only on the spectral space $\mathbb{CP}^{1}$. The combined action of both, in a way to be defined below, pervades the gauge fixing procedure and manifests itself in determining the explicit form of the R-matrix as well as other important quantities, like the Lax connection.

There is only one major section \eqref{2} containing all the details of the derivation of \eqref{1} in the symmetric space case. At the end, we comment on some open problems to be considered in the future.

\section{Holomorphic Chern-Simons theory}\label{2}

In this section, we introduce the 4d CS theory and a group action whose main role is to induce the coset structure on the resulting integrable lambda model that is found after gauge fixing. It also determines the explicit form of the extended Lax connection and the R-matrix in the exchange algebra. To see how this occurs, we run the Hamiltonian analysis of the CS theory by endowing it with an appropriate covariant Poisson bracket.

\subsection{Action and extended Lax connection}

The holomorphic Chern-Simons theory is defined by the 4d action functional \cite{CY},
\begin{equation}
S_{CS}=ic\int\nolimits_{\Sigma \times \mathbb{CP}^{1}}\omega \wedge CS(B),\qquad CS(B)=\big\langle B\wedge \hat{d}B+\frac{2}{3}B\wedge B\wedge B\big\rangle,\label{Hol CS action}
\end{equation} 
where $CS(B)$ is the CS 3-form for the gauge field $B$, $\Sigma=\mathbb{R}\times S^{1}$ is the closed string world-sheet manifold parameterized by $(\tau,\sigma)$, $\mathbb{CP}^{1}$ is the spectral space with local holomorphic coordinate $z$, $\omega=\varphi(z) dz$ is a meromorphic $(1,0)$-form to be defined below and $c$ is a constant.

The underlying Lie algebra $\mathfrak{f}$ of the CS theory is endowed with a $\mathbb{Z}_{4}$ decomposition induced by the automorphism $\Phi $,
\begin{equation}
\Phi \mathfrak{f}^{(a)}=i^{a}\mathfrak{f}^{(a)},\qquad\mathfrak{f=\;}%
\mathfrak{f}^{(0)}\oplus \mathfrak{f}^{(2)},\qquad[\mathfrak{f}%
^{(a)},\mathfrak{f}^{(b)}]\subset \mathfrak{f}^{(a+b)\func{mod}4},  \label{auto}
\end{equation}%
where\footnote{When $\mathfrak{f}$ is a Lie superalgebra, $a=0,1,2,3$, with $a=1,3$ corresponding to the fermionic sector. In this note, we ignore that sector but maintain the notation in order to facilitate its reintroduction.} $a,b=0,2$. Denote by $F$ and $G$ the Lie groups associated to $\mathfrak{f}$ and $\mathfrak{f}^{(0)}$, respectively, and recall that
\begin{equation}
\big\langle \mathfrak{f}^{(a)},\mathfrak{f}^{(b)}\big\rangle \neq 0,
\end{equation}%
when $\left( a+b\right) \func{mod}4=0$ and where $\big\langle \mathfrak{f}^{(a)},\mathfrak{f}^{(b)}\big\rangle$ is a non-degenerate inner product.\\ 
There is an action, on the spectral space $\mathbb{CP}^{1}$, of the $\mathbb{Z}_{4}$ cyclic group generated by the element $\rho=i$, i.e. $\rho^{4}=1$, as $z\rightarrow z'=iz$. \\
Also notice that, because of $\Phi^{4}=1$, the first expressions in \eqref{auto} takes the form
\begin{equation}
\Phi \mathfrak{f}^{(a)}=\rho^{a}\mathfrak{f}^{(a)}. 
\end{equation}
This allows to consider a single action of the $\mathbb{Z}_{4}$ cyclic group and to introduce an important equivariant map to be defined right below.

Define $x=(\tau,\sigma,z,\overline{z})$, $\tilde{x}=(\tau,\sigma,z)$, extend trivially the action of $\rho$ in the form
\begin{equation}
\rho\cdot x= (\tau,\sigma, iz,-i\overline{z}),\qquad \rho\cdot \tilde{x}=(\tau,\sigma, iz)
\end{equation}
and consider the map defined by
\begin{equation}
\psi: x\longrightarrow B=B_{j}(\tilde{x})dx^{j},
\end{equation}
with $j=\tau,\sigma,z,\overline{z}$, from the set $x$ to the CS 1-form $B$. \\
In the following, we will restrict the map $\psi$ to be equivariant under the action of the $\mathbb{Z}_{4}$ cyclic group, in the sense that $\psi \circ \rho=\Phi \circ \psi$, i.e.
\begin{equation}
\psi(\rho\cdot x)=\Phi\cdot \psi(x),
\end{equation} 
or, equivalently,
\begin{equation}
B_{j}(\rho\cdot \tilde{x})d(\rho\cdot x^{j})=\Phi B_{j}(\tilde{x})dx^{j}. \label{the-symmetry}
\end{equation}
We also demand invariance of the action functional under the action of the $\mathbb{Z}_{4}$ cyclic group. As a consequence of this, the resulting integrable string sigma model background will be a coset of $F$ by $G$. Below, we will discuss, from the Hamiltonian theory point of view, the compatibility between the action of the proper gauge group and the equivariance condition defined right above. 

The field $B$ and the exterior derivative $\hat{d}$ decompose in the form
\begin{equation}
\begin{aligned}
B &=A_{\tau }d\tau +A,\qquad \: \, A=A_{\sigma }d\sigma +A_{\overline{z}}d%
\overline{z}, \\
\hat{d} &=d\tau \wedge \partial _{\tau }+d,\qquad d=d\sigma \wedge
\partial _{\sigma }+dz\wedge \partial _{z}+d\overline{z}\wedge \partial _{%
\overline{z}},
\end{aligned} \label{quantities}
\end{equation}%
where we have ignored the $A_{z}dz$ component as it decouples from the theory. Explicitly, the condition \eqref{the-symmetry}, in terms of the field components, is\footnote{Explicitly, the fields gain a phase factor after a counterclockwise rotation by an angle of $\theta=\pi/2$, i.e. $A^{(a)}_{\mu}(\rho\cdot \tilde{x})=e^{ia\pi/2}A_{\mu}^{(a)}(\tilde{x})$ and $A^{(a)}_{\overline{z}}(\rho\cdot \tilde{x})=e^{i(a+1)\pi/2}A_{\overline{z}}^{(a)}(\tilde{x})$.}
\begin{equation}
A_{\mu }(\tau, \sigma, iz) =\Phi A_{\mu }(\tau, \sigma,z), \qquad A_{\overline{z}}(\tau, \sigma,iz) =i\Phi A_{\overline{z}}(\tau, \sigma,z), \label{comp-trans}
\end{equation}
where $\mu=\tau ,\sigma$. \\
A comment is in order. When we take $\mu=\sigma$ right above, we get a particular case of the more general $\mathbb{Z}_{T}$-equivariance condition\footnote{I would like to thank B. Vicedo for pointing me out this condition in an early discussion.} imposed on the component field $A_{\sigma}(\sigma,z)$ first considered in \cite{V3} (see also the discussion section of \cite{Vicedo-Holo}) and corresponding to the family of $\mathbb{Z}_{T}$-cyclotomic affine Gaudin models. The latter being related to (semi)-symmetric space sigma models, in the language of \cite{V3}. In this way, the condition \eqref{the-symmetry} is an extension of such an equivariance condition to include all the components of the 1-form gauge field $B$ of the CS theory.  

In the variables \eqref{quantities}, the action functional becomes
\begin{equation}
S_{CS}=ic\int\nolimits_{\Sigma \times \mathbb{CP}^{1}}d\tau \wedge \omega \wedge \big\langle A\wedge \partial _{\tau
}A-2A_{\tau }F\big\rangle +ic\int\nolimits_{\Sigma \times\mathbb{CP}^{1}}d\tau \wedge d\omega \wedge \big\langle A_{\tau }A\big\rangle  \label{Hol-CS-decomp}
\end{equation}%
and it is invariant under the action of the $\mathbb{Z}_{4}$ cyclic group, provided we have $\varphi(iz)=-i\varphi(z)$.\\
An explicit solution to the latter condition is given by the twist function of the lambda deformed string sigma model in a symmetric space $F/G$ \cite{k-def},
\begin{equation}
\varphi(z)=\frac{az^{3}}{(z^{4}-z_{+}^{4})(z^{4}-z_{-}^{4})},\qquad z_{\pm}=\lambda^{\pm1/2},
\end{equation}
where $\lambda$ (usually taken to be $0<\lambda<1$) is the deformation parameter and $a$ a constant (not to be confused with the same letter $a=0,2$ denoting the gradings of the Lie algebra $\mathfrak{f}$). The poles $\mathfrak{p}$ and zeroes $\mathfrak{z}$ of the meromorphic differential $\omega=\varphi(z)dz$ specify the associated integrable field theory. They are located at
\begin{equation}
\mathfrak{p}=(\pm z_{+},\pm iz_{+},\pm z_{-},\pm iz_{-}),\qquad \mathfrak{z}=(0,\infty). \label{poles and zeroes}
\end{equation}
The eight poles are simple and the two zeroes are of order three. All poles are related by the action of $\rho$. 

Before we proceed, we derive an useful formula to compute integrals of the form
\begin{equation}
\int\nolimits_{\mathbb{CP}^{1}}d\omega F(z),
\end{equation}
for a function $F(z)$ taking values in $\mathbb{CP}^{1}$.\\
By expanding $\omega$ locally around the poles\footnote{We use the same letter $z$ for the local coordinate around each pole.}, as follows
\begin{equation}
\omega =\frac{1}{4}\frac{a}{(z_{+}^{4}-z_{-}^{4})}\dsum\limits_{\alpha
=0}^{3}\left( \frac{dz}{z-i^{\alpha }z_{+}}-\frac{dz}{z-i^{\alpha }z_{-}}%
\right), 
\end{equation}%
we find\footnote{We have used $\delta _{zz^{\prime }}=\delta (z-z^{\prime })=-\frac{1}{2\pi i}\partial _{%
\overline{z}}\frac{1}{z-z^{\prime }}$.}%
\begin{equation}
d\omega =r\dsum\limits_{\alpha
=0}^{3}\big\{ \delta (z-i^{\alpha }z_{+})-\delta (z-i^{\alpha
}z_{-})\big\} dz\wedge d\overline{z},
\end{equation}
where
\begin{equation}
r=\frac{i\pi }{2}\frac{a}{(z_{+}^{4}-z_{-}^{4})}.
\end{equation}
From this, we get
\begin{equation}
\int\nolimits_{\mathbb{CP}^{1}}d\omega F(z)=r\dsum\limits_{\alpha
=0}^{3}\left\{ F(i^{\alpha }z_{+})-F(i^{\alpha
}z_{-})\right\}. \label{residue for}
\end{equation}

Two useful compact notations to be used below are, the $\mathbb{Z}_{4}$ invariant combination of Dirac delta distributions 
\begin{equation}
\hat{\delta }_{zz^{\prime }}=\dsum\limits_{\alpha
=0}^{3} \delta (z-i^{\alpha }z^{\prime})
\end{equation}%
and the $S^{1} \times \mathbb{CP}^{1}$ spatial integral of the form
\begin{equation}
\big\langle X,Y\big\rangle_{(\sigma,z)}=\int\nolimits_{S^{1} \times\mathbb{CP}^{1}}d\sigma\wedge dz\wedge d\overline{z}\big\langle XY\big\rangle.
\end{equation}

Now, we consider the Lagrangian of the theory, which is given by
\begin{equation}
L=ic\int\nolimits_{S^{1} \times\mathbb{CP}^{1}}\omega \wedge \big\langle A\wedge \partial _{\tau
}A-2A_{\tau }F\big\rangle +ic\int\nolimits_{S^{1} \times\mathbb{CP}^{1}}d\omega \wedge \big\langle A_{\tau }A\big\rangle 
\end{equation}%
and has the following arbitrary variation 
\begin{equation}
\begin{aligned}
\delta L= 2ic\int\nolimits_{S^{1} \times \mathbb{CP}^{1}} \omega \wedge \big\langle \delta A \wedge (\partial_{\tau }A-DA_{\tau })-\delta A_{\tau }F\big\rangle +ic
\int\nolimits_{S^{1} \times \mathbb{CP}^{1}}d\omega \wedge \big\langle \delta A_{\tau }A-\delta AA_{\tau
}\big\rangle , \label{variation1}
\end{aligned}
\end{equation}%
where $D(\ast )=d(\ast )+[A,\ast ]$ and $F=dA+A\wedge A$.

By taking into account \eqref{comp-trans}, the last contribution to the variation \eqref{variation1} above, can be written in the form
\begin{equation}
ic\int\nolimits_{S^{1} \times \mathbb{CP}^{1}}d\omega \wedge \big\langle \delta A_{\tau }A-\delta A A_{\tau
}\big\rangle =-ic\big\langle \delta A_{\tau },\partial _{%
\overline{z}}( \varphi\mathscr{L}_{\sigma }) -\delta
A_{\sigma },\partial _{\overline{z}}( \varphi \mathscr{L}_{\tau
}) \big\rangle_{(\sigma,z)} , \label{Bdry term}
\end{equation}
where
\begin{equation}
\mathscr{L}_{\mu }(z)=f_{-}(z)\Omega (z/z_{+})A_{\mu }(z_{+})+f_{+}(z)\Omega
(z/z_{-})A_{\mu }(z_{-})\label{extended-Lax}
\end{equation}%
and
\begin{equation}
\Omega (z)=P^{(0)}+z^{-2}P^{(2)},\qquad f_{\pm }(z)=\mp \frac{%
(z^{4}-z_{\pm }^{4})}{(z_{+}^{4}-z_{-}^{4})}.
\end{equation}
The $P^{(a)}$, with $a=0,2$ are projectors along the graded spaces of the Lie algebra $\mathfrak{f}$.\\
The expression \eqref{extended-Lax} is recognized as having the same structure of the extended Lax connection \cite{lambdaCS,lambdaCS2} of the lambda model. Two properties of \eqref{extended-Lax} are that it interpolates between $A_{\mu}(z_{+})$ and $A_{\mu}(z_{-})$ and manifests the equivariance condition \eqref{comp-trans} explicitly, i.e.
\begin{equation}
\mathscr{L}_{\mu}(iz)=\Phi \mathscr{L}_{\mu}(z).
\end{equation}
Because of \eqref{the-symmetry}, we have that $\mathscr{L_{\mu}}(z)\in \hat{\mathfrak{f}}$ actually is valued in the twisted loop algebra $\hat{\mathfrak{f}}$ defined below in \eqref{loop algebra}. Later on, we will come back to \eqref{extended-Lax} and comment on its meaning with more detail. 

The term \eqref{Bdry term} is a `boundary' contribution to the Euler-Lagrange variational problem. In order to cancel it, we take the following ansatz for the time component
\begin{equation}
A_{\tau}(z_{\pm})=a_{+}(z_{\pm})A_{\sigma}^{(0)}(z_{+})+b_{+}(z_{\pm})A_{\sigma}^{(2)}(z_{+})+a_{-}(z_{\pm})A_{\sigma}^{(0)}(z_{-})+b_{-}(z_{\pm})A_{\sigma}^{(2)}(z_{-}), \label{Ansats for Atau}
\end{equation} 
where $a_{+}(z_{\pm}),b_{+}(z_{\pm}),a_{-}(z_{\pm}),b_{-}(z_{\pm})$ are constants to be determined. The vanishing of the `boundary' term, then requires that
\begin{equation}
a_{-}(z_{+})=-a_{+}(z_{-}),\qquad b_{-}(z_{+})=-b_{+}(z_{-}). \label{vanish bdry}
\end{equation}
The precise values for all constants are unique up to a sign and will be obtained later from the Hamiltonian analysis. The only condition imposed over the constants, at this stage, is \eqref{vanish bdry}. \\
The ansatz \eqref{Ansats for Atau} deserves a comment. In \eqref{Ansats for Atau} we are anticipating that, after gauge fixing the CS theory, the resulting reduced phase space variables are given by the $\sigma$-component of the CS gauge field localized at the set of poles\footnote{An example of this type of localization was shown to occur in the lambda deformed PCM, after symplectic reduction \cite{PCM-Case}. This is the motivation behind \eqref{Ansats for Atau}. } $\mathfrak{p}$.   

The equations of motion (eom) derived from \eqref{variation1}, are given by%
\begin{equation}
\varphi F_{\tau \sigma } =0, \qquad \varphi F_{\overline{z}\mu } =0. \label{full eom}
\end{equation}%
By restricting (by hand) the last pair of eom (for $\mu=\tau,\sigma$) to field configurations where $A_{\overline{z}}=0$,
we find that
\begin{equation}
\partial_{\overline{z}}(z^{3}A_{\mu }(z))=0. \label{condition ove Lax}
\end{equation}%
A solution to \eqref{condition ove Lax} is provided by \eqref{extended-Lax}, which has the right $1/z$ dependence and also suggests to take the gauge fixing condition $A_{\overline{z}}\approx 0$ in the Hamiltonian formalism.\\
The remaining eom
\begin{equation}
\varphi F_{\tau \sigma }=0, \label{last eom}
\end{equation}
turns out to be equivalent to the extended eom of the symmetric space lambda model \cite{lambdaCS2}, in the sense of being a strong flatness condition for the extended Lax connection. More on this below. The outcome is that, after gauge fixing the CS theory, the integrability of the reduced theory is described via \eqref{extended-Lax} in terms of the phase space coordinates $A_{\sigma}(z_{\pm})$ endowed with a Poisson bracket of the Kac-Moody type.

In order to give a more solid ground to these observations, we consider the 4d CS theory from the Hamiltonian theory point of view.

\subsection{Hamiltonian analysis}

The canonical Hamiltonian of the theory is given by
\begin{equation}
h=2ic\dint\nolimits_{S^{1}\times \mathbb{CP}^{1}}\omega \wedge \big\langle A_{\tau }F\big\rangle -ic
\dint\nolimits_{S^{1}\times \mathbb{CP}^{1}}d\omega \wedge \big\langle A_{\tau }A\big\rangle  \label{can Ham CS}
\end{equation}%
and has a variation of the form
\begin{equation}
\begin{aligned}
\delta h= 2ic\int\nolimits_{S^{1} \times \mathbb{CP}^{1}} \omega \wedge \big\langle\delta A_{\tau }F+\delta A \wedge DA_{\tau }\big\rangle, \label{variation}
\end{aligned}
\end{equation}%
where we have canceled the `boundary' term contribution. In terms of the components, it is
\begin{equation}
\delta h=2ic\big\langle \delta A_{%
\overline{z}},\left\{ \varphi D_{\sigma }A_{\tau}\right\} +\delta A_{\sigma
},\left\{ -\varphi D_{\overline{z}}A_{\tau}\right\} +\delta A_{\tau },\left\{
\varphi F_{\overline{z}\sigma }\right\} \big\rangle_{(\sigma,z)} .
\end{equation}%

The Lagrangian is already in the first-order form
\begin{equation}
L=ic\big\langle \varphi \left( A_{\overline{z}},\partial _{\tau }A_{\sigma
}-A_{\sigma },\partial _{\tau }A_{\overline{z}}\right) \big\rangle_{(\sigma, z)} -h.
\end{equation}%
From this, we identify three primary constraints%
\begin{equation}
P_{\tau }\approx 0,\qquad \phi _{\sigma }=P_{_{\sigma }}-ic\varphi A_{_{%
\overline{z}}}\approx 0,\qquad \phi _{_{\overline{z}}}=P_{_{\overline{z}%
}}+ic\varphi A_{\sigma }\approx 0,\label{primary}
\end{equation}%
where $P_{j},$ $j=\tau ,\sigma ,\overline{z}$ is the conjugate momentum of $%
A_{j},$ $j=\tau ,\sigma ,\overline{z}$. From the conditions \eqref{comp-trans}, we obtain the equivariance conditions for the conjugate momentum, that make the constraints to have a well define behavior under the action of the $\mathbb{Z}_{4}$ cyclic group, namely
\begin{equation}
P_{\mu }(\tau ,\sigma ,iz) =\Phi P_{\mu }(\tau ,\sigma ,z), \qquad P_{\overline{z}}(\tau ,\sigma ,iz) =-i\Phi P_{\overline{z}}(\tau ,\sigma
,z). \label{comp-trans P}
\end{equation}

The theory is endowed with the following canonical Poisson bracket%
\begin{equation}
\left\{ f,g\right\} =\frac{1}{4}\left\langle 
\begin{array}{c}
\frac{\delta f}{\delta A_{\mu }(\sigma ,z)},\Phi ^{\alpha }\left(\frac{%
\delta g}{\delta P_{\mu }(\sigma ,i^{\alpha }z)}\right) -\frac{\delta f}{%
\delta P_{\mu }(\sigma ,z)},\Phi ^{\alpha }\left( \frac{\delta g}{\delta
A_{\mu }(\sigma ,i^{\alpha }z)}\right)  \\ 
+\frac{\delta f}{\delta A_{\overline{z}}(\sigma ,z)},\left( \frac{\Phi }{i}%
\right) ^{\alpha }\left( \frac{\delta g}{\delta P_{\overline{z}}(\sigma
,i^{\alpha }z)}\right) -\frac{\delta f}{\delta P_{\overline{z}}(\sigma ,z)}%
,\left( i\Phi \right) ^{\alpha } \left( \frac{\delta g}{\delta A_{\overline{z}%
}(\sigma ,i^{\alpha }z)}\right) 
\end{array}%
\right\rangle _{(\sigma ,z)}, \label{CS-PB}
\end{equation}
where sum over $\mu=\tau,\sigma$ and $\alpha=0,1,2,3$ is implied. This Poisson bracket reduce to the standard one when functionals are scalars under the action of the $\mathbb{Z}_{4}$ cyclic group, but provide covariant expressions for other kind of functionals, in the sense of being compatible with the conditions \eqref{the-symmetry} and \eqref{comp-trans P}.

To see how to operate with \eqref{CS-PB}, let us compute a sample bracket. Consider, for example,
\begin{equation}
f=A_{\overline{z}}^{(2)}(\sigma,z), \qquad g=P_{\overline{z}}^{(2)}(\sigma^{\prime},z').
\end{equation}
Then, we have that 
\begin{equation}
\left\{ A_{\overline{z}}^{(2)}(\sigma ,z)_{\mathbf{1}},P_{\overline{z}}^{(2)}(\sigma ^{\prime
},z^{\prime })_{\mathbf{2}}\right\}=\frac{1}{4}\left\langle 
\begin{array}{c}
\frac{\delta A_{\overline{z}}^{(2)}(\sigma ,z)_{\mathbf{1}}}{\delta A_{\overline{z}}^{(2)}(\sigma'' ,z'')_{\mathbf{3}}},\left( \frac{\Phi }{i}
\right) ^{\alpha }_{\mathbf{3}}\left( \frac{\delta P_{\overline{z}}^{(2)}(\sigma'
,z')_{\mathbf{2}}}{\delta P_{\overline{z}}^{(2)}(\sigma''
,i^{\alpha }z'')_{\mathbf{3}}}\right) 
\end{array}%
\right\rangle _{(\sigma'' ,z'')\mathbf{3}},
\end{equation}
where the extra index $\mathbf{3}$ in $(\sigma'',z'')\mathbf{3}$ instructs to take the trace along the corresponding tensor factor. Now, consider the functional derivatives\footnote{We use $\eta_{AB}=\langle T_{A},T_{B} \rangle$, the tensor Casimir $C_{\mathbf{12}}=\eta^{AB}T_{A}\otimes T_{B}$ and $\delta_{\sigma \sigma'}=\delta(\sigma-\sigma')$.}
\begin{equation}
\frac{\delta A_{\overline{z}}^{(2)}(\sigma ,z)_{\mathbf{1}}}{\delta A_{%
\overline{z}}^{(2)}(\sigma ^{\prime \prime },z^{\prime \prime })_{\mathbf{3}}%
}=C_{\mathbf{13}}^{(22)}\delta _{\sigma \sigma ^{\prime \prime }}\delta
_{zz^{\prime \prime }},\qquad \frac{\delta P_{\overline{z}}^{(2)}(\sigma ^{\prime
},z^{\prime })_{\mathbf{2}}}{\delta P_{\overline{z}}^{(2)}(\sigma ^{\prime
\prime },z^{\prime \prime })_{\mathbf{3}}}=C_{\mathbf{23}}^{(22)}\delta
_{\sigma ^{\prime }\sigma ^{\prime \prime }}\delta _{z^{\prime }z^{\prime
\prime }}
\end{equation}%
and the fact that $\Phi_{\mathbf{3}}C_{\mathbf{23}}^{(22)}=i^{2}C_{\mathbf{23}}^{(22)}$. Then,
\begin{equation}
\left( \frac{\Phi }{i}\right) _{\mathbf{3}}^{\alpha }\frac{\delta P_{%
\overline{z}}^{(2)}(\sigma ^{\prime },z^{\prime })_{\mathbf{2}}}{\delta P_{%
\overline{z}}^{(2)}(\sigma ^{\prime \prime },i^{\alpha }z^{\prime \prime })_{%
\mathbf{3}}}=C_{\mathbf{23}}^{(22)}\delta _{\sigma ^{\prime }\sigma ^{\prime
\prime }}i^{\alpha }\delta (z'-i^{\alpha }z^{\prime \prime })=C_{\mathbf{23}%
}^{(22)}\delta _{\sigma ^{\prime }\sigma ^{\prime \prime }}\frac{z^{\prime }%
}{z^{\prime \prime }}\hat{\delta }_{z^{\prime }z^{\prime \prime }}.
\end{equation}%
In this way, 
\begin{equation}
\begin{aligned}
\left\{ A_{\overline{z}}^{(2)}(\sigma ,z)_{\mathbf{1}},P_{\overline{z}}^{(2)}(\sigma ^{\prime
},z^{\prime })_{\mathbf{2}}\right\}&=\frac{1}{4}\left\langle 
\begin{array}{c}
C_{\mathbf{13}}^{(22)}\delta _{\sigma \sigma ^{\prime \prime }}\delta
_{zz^{\prime \prime }},C_{\mathbf{23}%
}^{(22)}\delta _{\sigma ^{\prime }\sigma ^{\prime \prime }}\frac{z^{\prime }%
}{z^{\prime \prime }}\hat{\delta }_{z^{\prime }z^{\prime \prime }} 
\end{array}%
\right\rangle _{(\sigma'' ,z'')\mathbf{3}}\\
&=\frac{1}{4}C_{\mathbf{12}%
}^{(22)}\delta _{\sigma \sigma ^{\prime }} \frac{z^{\prime }}{z}%
\hat{\delta }_{zz^{\prime }}.
\end{aligned}
\end{equation}

After repeating a similar calculation for all the component fields, we get the following covariant canonical Poisson brackets
\begin{equation}
\begin{aligned}
\left\{ A_{\mu }^{(0)}(\sigma ,z)_{\mathbf{1}},P_{\nu }^{(0)}(\sigma
^{\prime },z^{\prime })_{\mathbf{2}}\right\}  &=\frac{1}{4}C_{\mathbf{12}%
}^{(00)}\delta_{\mu\nu}\delta _{\sigma \sigma ^{\prime }}\hat{\delta }_{zz^{\prime }},
\\
\left\{ A_{\mu }^{(2)}(\sigma ,z)_{\mathbf{1}},P_{\nu }^{(2)}(\sigma
^{\prime },z^{\prime })_{\mathbf{2}}\right\}  &=\frac{1}{4}C_{\mathbf{12}%
}^{(22)}\delta_{\mu\nu}\delta _{\sigma \sigma ^{\prime }}\left( \frac{z}{z^{\prime }}%
\right) ^{2s}\hat{\delta }_{zz^{\prime }}, \\
\left\{ A_{_{\overline{z}}}^{(0)}(\sigma ,z)_{\mathbf{1}},P_{_{\overline{z}%
}}^{(0)}(\sigma ^{\prime },z^{\prime })_{\mathbf{2}}\right\}  &=\frac{1}{4}%
C_{\mathbf{12}}^{(00)}\delta _{\sigma \sigma ^{\prime }}\frac{z}{z^{\prime }}%
\hat{\delta }_{zz^{\prime }}, \\
\left\{ A_{\overline{z}}^{(2)}(\sigma ,z)_{\mathbf{1}},P_{\overline{z}}^{(2)}(\sigma ^{\prime
},z^{\prime })_{\mathbf{2}}\right\}  &=\frac{1}{4}C_{\mathbf{12}%
}^{(22)}\delta _{\sigma \sigma ^{\prime }} \frac{z^{\prime }}{z}%
\hat{\delta }_{zz^{\prime }}.
\end{aligned} \label{canonical PB}
\end{equation}
The choice of $s=\pm 1$ being arbitrary. Both sides are consistent under the conditions \eqref{comp-trans} and \eqref{comp-trans P}. We call them covariant brackets because they are devised to manifest the equivariance conditions of the fields under the action of the $\mathbb{Z}_{4}$ cyclic group $\rho \cdot z=iz$.

The total Hamiltonian takes the form
\begin{equation}
h_{T}=h+\big\langle u_{\tau },P_{\tau }+u_{\sigma },\phi _{\sigma }+u_{%
\overline{z}},\phi _{\overline{z}}\big\rangle
_{(\sigma ,z)} \label{total Hamiltonian}
\end{equation}%
and it is a scalar provided the Lagrange multipliers satisfy the conditions
\begin{equation}
u_{\mu}(\tau,\sigma,iz)=\Phi u_{\mu}(\tau,\sigma,z)\qquad u_{\overline{z}}(\tau,\sigma,iz)=i\Phi u_{\overline{z}}(\tau,\sigma,z), 
\end{equation}
which we assume to be the case.

We are now ready to verify the time preservation of the primary constraints. Using the variations 
\begin{equation}
\frac{\delta h_{T}}{\delta A_{\tau }}=\frac{\delta h}{\delta A_{\tau }},%
\qquad\frac{\delta h_{T}}{\delta A_{\sigma }}=\frac{\delta h}{%
\delta A_{\sigma }}+ic\varphi u_{\overline{z}},\qquad%
\frac{\delta h_{T}}{\delta A_{\overline{z}}}=\frac{\delta h}{\delta A_{%
\overline{z}}}-ic\varphi u_{\sigma },
\end{equation}%
with%
\begin{equation}
\frac{\delta h}{\delta A_{\tau }}=2ic \varphi F_{\overline{z}%
\sigma } ,\qquad\frac{\delta h}{\delta
A_{\sigma }}=-2ic \varphi D_{\overline{z}}A_{\tau } ,\qquad\frac{\delta h}{\delta A_{\overline{z}}}%
=2ic\varphi D_{\sigma }A_{\tau },
\end{equation}%
we find that
\begin{equation}
\begin{aligned}
\left\{ h_{T},P_{\tau }\right\}  &=2ic\varphi F_{\overline{z}\sigma
} \approx 0, \\
\left\{ h_{T},\phi _{\sigma }\right\}  &= \frac{\delta h}{\delta
A_{\sigma }}+2ic\varphi u_{\overline{z}} \approx 0, \\
\left\{ h_{T},\phi _{_{\overline{z}}}\right\}  &= \frac{%
\delta h}{\delta A_{\overline{z}}}-2ic\varphi u_{\sigma } \approx
0.\label{invariant ex}
\end{aligned}
\end{equation}%
The last two conditions determine two of the Lagrange multipliers to be%
\begin{equation}
u_{\sigma }\approx D_{\sigma }A_{\tau },\qquad u_{\overline{z}}\approx D_{\overline{z}}A_{\tau },\label{us}
\end{equation}%
while the first condition provide a secondary constraint%
\begin{equation}
\gamma =2ic \varphi F_{\overline{z}\sigma }\approx 0. \label{secondary}
\end{equation}

The expressions \eqref{invariant ex} are examples of the use of the canonical bracket \eqref{CS-PB}, when one of the functionals is a scalar. The total Hamiltonian in this case. Let us consider an example, say the first bracket, to see how \eqref{CS-PB} is used in such a situation. Taking,
\begin{equation}
f=P_{\tau}(\sigma,z),\qquad g=h_{T},
\end{equation}
we have that
\begin{equation}
\left\{ P_{\tau}(\sigma ,z),h_{T}\right\}=-\frac{1}{4}\left\langle 
\begin{array}{c}
\frac{\delta P_{\tau}(\sigma ,z)_{\mathbf{1}}}{\delta P_{\tau}(\sigma' ,z')_{\mathbf{2}}}, \Phi
 ^{\alpha }_{\mathbf{2}}\left( \frac{\delta h_{T}}{\delta A_{\tau}(\sigma'
,i^{\alpha }z')_{\mathbf{2}}}\right) 
\end{array}%
\right\rangle _{(\sigma' ,z')\mathbf{2}}.
\end{equation}
The functional variations are
\begin{equation}
\frac{\delta P_{\tau }(\sigma ,z)_{\mathbf{1}}}{\delta P_{\tau }(\sigma
^{\prime },z^{\prime })_{\mathbf{2}}}=C_{\mathbf{12}}\delta _{\sigma \sigma
^{\prime }}\delta _{zz^{\prime }},\qquad\frac{\delta h_{T}}{\delta
A_{\tau }(\sigma ^{\prime },z^{\prime })_{\mathbf{2}}}=2ic\varphi (z^{\prime
})F_{\overline{z}\sigma }(\sigma ^{\prime },z^{\prime })_{\mathbf{2}},
\end{equation}%
giving%
\begin{eqnarray}
\Phi_{\mathbf{2}}^{\alpha }\left( \frac{\delta h_{T}}{\delta
A_{\tau }(\sigma ^{\prime },i^{\alpha }z^{\prime })_{\mathbf{2}}}\right) 
&=&2ic\varphi (i^{\alpha }z^{\prime })\left( F_{\overline{z}\sigma
}^{(0)}(\sigma ^{\prime },i^{\alpha }z^{\prime })_{\mathbf{2}}+i^{2\alpha
}F_{\overline{z}\sigma }^{(2)}(\sigma ^{\prime },i^{\alpha }z^{\prime })_{%
\mathbf{2}}\right)  \\
&=&4\times 2ic\varphi (z^{\prime })F_{\overline{z}\sigma }(\sigma ^{\prime
},z^{\prime })_{\mathbf{2}},
\end{eqnarray}%
where we have used%
\begin{equation}
F_{\overline{z}\sigma }^{(0)}(i^{\alpha }z)=i^{\alpha }F_{\overline{z}\sigma
}^{(0)}(z),\qquad F_{\overline{z}\sigma }^{(2)}(i^{\alpha }z)=i^{-\alpha }F_{%
\overline{z}\sigma }^{(2)}(z),\qquad \varphi (i^{\alpha }z)=i^{-\alpha }\varphi (z).
\end{equation}
In this way,
\begin{equation}
\begin{aligned}
\left\{ P_{\tau}(\sigma ,z),h_{T}\right\}&=-\left\langle 
\begin{array}{c}
C_{\mathbf{12}}\delta _{\sigma \sigma
^{\prime }}\delta _{zz^{\prime }}, 2ic\varphi (z^{\prime })F_{\overline{z}\sigma }(\sigma ^{\prime
},z^{\prime })_{\mathbf{2}} 
\end{array}%
\right\rangle _{(\sigma' ,z')\mathbf{2}}.\\
&=-2ic\varphi(z)F_{\overline{z}\sigma }(\sigma ,z).
\end{aligned}
\end{equation}
The other brackets are computed in a similar fashion. The factor of $1/4$ in \eqref{CS-PB} is there precisely to guarantee the correct result when, either $f$ or $g$ is a scalar or both are scalars. 

Now return to the Hamiltonian analysis. Before we evaluate the time evolution of the secondary constraint, consider the functional
\begin{equation}
G(\eta)=2ic\dint\nolimits_{S^{1}\times \mathbb{CP}^{1}}\omega\wedge\big\langle\eta F\big\rangle-2ic\dint\nolimits_{S^{1}\times \mathbb{CP}^{1}}d\omega\wedge\big\langle\eta A\big\rangle.
\end{equation}
It is a scalar for gauge parameters $\eta$ satisfying the conditions
\begin{equation}
\eta(\tau,\sigma,iz)=\Phi \eta(\tau,\sigma,z) \label{gauge par equiv}
\end{equation} 
and has a well-defined functional variation, in the sense that no `boundary' terms are generated.
The treatment of boundary terms in the Hamiltonian framework was first addressed in the seminal paper \cite{RT} in the context of general relativity,
see also \cite{BH1,BH2}. The application of these ideas to ordinary Chern-Simons theory,
directly relevant to the present discussion, was considered in \cite{B1,B2}, see also \cite{BR}. \\
Using,%
\begin{equation}
\frac{\delta G(\eta )}{\delta A_{\overline{z}}}=2ic\varphi D_{\sigma }\eta ,%
\qquad\frac{\delta G(\eta )}{\delta A_{\sigma }}=-2ic\varphi D_{%
\overline{z}}\eta,
\end{equation}%
we find that%
\begin{equation}
\left\{ h_{T},G(\eta )\right\} =-2ic\dint\nolimits_{S^{1}\times \mathbb{CP}^{1}}\omega \wedge \left\langle u\wedge D\eta \right\rangle,
\end{equation}%
where we have defined%
\begin{equation}
u=u_{\sigma }d\sigma +u_{\overline{z}}d\overline{z}.
\end{equation}
Now, replacing \eqref{us} and imposing \eqref{secondary}, we get%
\begin{equation}
\left\{ h_{T},G(\eta )\right\} \approx 2ic\dint\nolimits_{S^{1}\times 
\mathbb{CP}^{1}}d\omega \wedge \left\langle \eta DA_{\tau}\right\rangle \approx 0.
\end{equation}%
The secondary constraint is preserved in time when $\eta $ vanish at the
poles \eqref{poles and zeroes} and when this occurs, we denote such a constraint by $G_{0}(\eta )$. No tertiary constraints are produced. 

Summarizing, we have three primary constraints and one secondary constraint
given, respectively, by 
\begin{equation}
P_{\tau }\approx 0,\qquad\phi _{\sigma }\approx 0,\qquad\phi _{\overline{z}}\approx 0,\qquad G_{0}(\eta )\approx 0.
\end{equation}%
The time evolution and Poisson brackets, at this stage, are given by \eqref{total Hamiltonian} and \eqref{canonical PB}.

Before we catalog the constraints as first class or second class, we
realize that $\phi _{\sigma }$ and $\phi _{\overline{z}}$
form a pair of second class constraints and we now proceed to impose them
strongly by means of a Dirac bracket. We do this by sectors.    

\subsubsection{Emergence of the R-matrix}

First, the Poisson bracket for $\mu=\tau$ in \eqref{canonical PB} is not modified. Now, we consider the different Lie algebra gradings separately. The main result at this stage is the emergence of the symmetric space lambda model R-matrix.

\textbf{Grade zero sector.}\ 

We need to compute the Dirac bracket\footnote{Recall that $\mathbf{3}$ and $\mathbf{4}$ indicate taking the trace.}
\begin{multline}
\! \! \! \! \! \! \big\{A_{\overline{z}}^{(0)}(\sigma,z)_{\mathbf{1}},P_{\overline{z}}^{(0)}(\sigma^{\prime},z^{\prime})_{\mathbf{2}}\big\}^{\ast}=\big\{A_{\overline{z}}^{(0)}(\sigma,z)_{\mathbf{1}},P_{\overline{z}}^{(0)}(\sigma^{\prime},z^{\prime})_{\mathbf{2}}\big\}\\
\! \! \! \! \! \! \! \! \! \! \! \! \! \! \! \! \! \! \! \! \! \! \! \! \! \! \! \! \! \! \! \! \! \! \! \! \! \! \! \! \! \! \! \! \! \! \! \! 
\! \! \! \! \! \! \! \! \! \! \! \! \! \! \! \! \! \! \! \! \! \! \! \! \! \! \! \! \! \! \! \! \! \! \! \! \! \! \! \! \! \! \! \! \! \! \! \! 
\! \! \! \! \! \! \! \! \! \! \! \! \! \! \! \! \! \! \! \! \! \! \! \! \! \! \! \! \! \!
\! \! \! \! \! \! \! \! \! \! \! \! \! \! \! \! \! \! \! \! \! \! \! \! \! \! \! \! \! \! \! \!
-\bigg\langle \! \left\{ A_{\overline{z}}^{(0)}(\sigma,z)_{\mathbf{1}},\phi_{\overline{z} }^{(0)}(\sigma
^{\prime \prime },z^{\prime \prime })_{\mathbf{3}}\right\} ,\\
\qquad \qquad \Big\langle
\! \left\{ \phi_{\overline{z} }^{(0)}(\sigma
^{\prime \prime },z^{\prime \prime })_{\mathbf{3}},\phi_{%
\sigma}^{(0)}(\sigma ^{\prime \prime \prime },z^{\prime \prime \prime })_{%
\mathbf{4}}\right\} ^{-1},\left\{ \phi_{\sigma}^{(0)}(\sigma ^{\prime \prime
\prime },z^{\prime \prime \prime })_{\mathbf{4}},P_{\overline{z}}^{(0)}(\sigma ^{\prime
},z^{\prime })_{\mathbf{2}}\right\}\! \Big\rangle _{(\sigma ^{\prime \prime
\prime },z^{\prime \prime \prime })\mathbf{4}}\bigg\rangle _{(\sigma
^{\prime \prime },z^{\prime \prime })\mathbf{3}}.
\end{multline}

From \eqref{primary}, the grade zero constraint algebra is
\begin{equation}
\left\{ \phi _{\sigma }^{(0)}(\sigma ,z)_{\mathbf{1}},\phi _{_{\overline{z}%
}}^{(0)}(\sigma ^{\prime },z^{\prime })_{\mathbf{2}}\right\} =-\frac{1}{2}%
ic\varphi (z^{\prime })C_{\mathbf{12}}^{(00)}\delta _{\sigma \sigma
^{\prime }}\hat{\delta }_{zz^{\prime }}.
\end{equation}%
Its inverse is defined by the condition
\begin{equation}
\begin{aligned}
& \left\langle \left\{ \phi _{\sigma }^{(0)}(\sigma ,z)_{\mathbf{1}},\phi _{_{%
\overline{z}}}^{(0)}(\sigma ^{\prime \prime },z^{\prime \prime })_{\mathbf{3}%
}\right\} ,\left\{ \phi _{_{\overline{z}}}^{(0)}(\sigma ^{\prime \prime
},z^{\prime \prime })_{\mathbf{3}},\phi _{\sigma }^{(0)}(\sigma ^{\prime
},z^{\prime })_{\mathbf{2}}\right\} ^{-1}\right\rangle_{(\sigma^{\prime \prime},z^{\prime \prime})\mathbf{3}}\\
&\quad  =C_{\mathbf{12}%
}^{(00)}\delta _{\sigma \sigma ^{\prime }}\hat{\delta }_{zz^{\prime }}
\end{aligned}
\end{equation}%
and it is covariant under \eqref{comp-trans} and \eqref{comp-trans P}. We find,%
\begin{equation}
\left\{ \phi _{_{\overline{z}}}^{(0)}(\sigma ,z)_{\mathbf{1}},\phi _{\sigma }^{(0)}(\sigma ^{\prime },z^{\prime })_{%
\mathbf{2}},\right\} ^{-1}=\frac{i}{2c}\varphi (z)^{-1}C_{\mathbf{12}%
}^{(00)}\delta _{\sigma \sigma ^{\prime }}\hat{\delta }_{zz^{\prime }}. \label{inverse 0}
\end{equation}%
From this result it follows that 
\begin{equation}
\begin{aligned}
& \left\langle
\! \left\{ \phi_{\overline{z} }^{(0)}(\sigma
^{\prime \prime },z^{\prime \prime })_{\mathbf{3}},\phi_{%
\sigma}^{(0)}(\sigma ^{\prime \prime \prime },z^{\prime \prime \prime })_{%
\mathbf{4}}\right\} ^{-1},\left\{ \phi_{\sigma}^{(0)}(\sigma ^{\prime \prime
\prime },z^{\prime \prime \prime })_{\mathbf{4}},P_{\overline{z}}^{(0)}(\sigma ^{\prime
},z^{\prime })_{\mathbf{2}}\right\}\! \right\rangle _{(\sigma ^{\prime \prime
\prime },z^{\prime \prime \prime })\mathbf{4}}\\
&\quad =\frac{1}{2}C_{\mathbf{32}}%
^{(00)} \delta_{\sigma^{\prime \prime}\sigma^{\prime}}\frac{z''}{z'}\hat{\delta }_{z''z^{\prime }}.
\end{aligned}
\end{equation}
Hence,
\begin{equation}
\big\{A_{\overline{z}}^{(0)}(\sigma,z)_{\mathbf{1}},P_{\overline{z}}^{(0)}(\sigma^{\prime},z^{\prime})_{\mathbf{2}}\big\}^{\ast}=\frac{1}{2}\big\{A_{\overline{z}}^{(0)}(\sigma,z)_{\mathbf{1}},P_{\overline{z}}^{(0)}(\sigma^{\prime},z^{\prime})_{\mathbf{2}}\big\}.
\end{equation}

In a similar way, we need to compute
\begin{multline}
\! \! \! \! \! \! \big\{A_{\sigma}^{(0)}(\sigma,z)_{\mathbf{1}},P_{\sigma}^{(0)}(\sigma^{\prime},z^{\prime})_{\mathbf{2}}\big\}^{\ast}=\big\{A_{\sigma}^{(0)}(\sigma,z)_{\mathbf{1}},P_{\sigma}^{(0)}(\sigma^{\prime},z^{\prime})_{\mathbf{2}}\big\}\\
\! \! \! \! \! \! \! \! \! \! \! \! \! \! \! \! \! \! \! \! \! \! \! \! \! \! \! \! \! \! \! \! \! \! \! \! \! \! \! \! \! \! \! \! \! \! \! \! 
\! \! \! \! \! \! \! \! \! \! \! \! \! \! \! \! \! \! \! \! \! \! \! \! \! \! \! \! \! \! \! \! \! \! \! \! \! \! \! \! \! \! \! \! \! \! \! \! 
\! \! \! \! \! \! \! \! \! \! \! \! \! \! \! \! \! \! \! \! \! \! \! \! \! \! \! \! \! \!
\! \! \! \! \! \! \! \! \! \! \! \! \! \! \! \! \! \! \! \! \! \! \! \! \! \! \! \! \! \! \! \!
-\bigg\langle \! \left\{ A_{\sigma}^{(0)}(\sigma,z)_{\mathbf{1}},\phi_{\sigma}^{(0)}(\sigma
^{\prime \prime },z^{\prime \prime })_{\mathbf{3}}\right\} ,\\
\qquad \qquad \Big\langle
\! \left\{ \phi_{\sigma }^{(0)}(\sigma
^{\prime \prime },z^{\prime \prime })_{\mathbf{3}},\phi_{%
\overline{z}}^{(0)}(\sigma ^{\prime \prime \prime },z^{\prime \prime \prime })_{%
\mathbf{4}}\right\} ^{-1},\left\{ \phi_{\overline{z}}^{(0)}(\sigma ^{\prime \prime
\prime },z^{\prime \prime \prime })_{\mathbf{4}},P_{\sigma}^{(0)}(\sigma ^{\prime
},z^{\prime })_{\mathbf{2}}\right\}\! \Big\rangle _{(\sigma ^{\prime \prime
\prime },z^{\prime \prime \prime })\mathbf{4}}\bigg\rangle _{(\sigma
^{\prime \prime },z^{\prime \prime })\mathbf{3}}.
\end{multline}
This time, the inverse is defined by the covariant condition
\begin{equation}
\begin{aligned}
& \left\langle \left\{ \phi _{_{\overline{z}}}^{(0)}(\sigma ,z)_{\mathbf{1}%
},\phi _{_{\sigma }}^{(0)}(\sigma ^{\prime \prime },z^{\prime \prime })_{%
\mathbf{3}}\right\} ,\left\{ \phi _{\sigma }^{(0)}(\sigma ^{\prime \prime
},z^{\prime \prime })_{\mathbf{3}},\phi _{_{\overline{z}}}^{(0)}(\sigma
^{\prime },z^{\prime })_{\mathbf{2}},\right\} ^{-1}\right\rangle _{(\sigma
^{\prime \prime },z^{\prime \prime })\mathbf{3}}\\
&\quad =C_{\mathbf{12}%
}^{(00)}\delta _{\sigma \sigma ^{\prime }}\frac{z^{\prime }}{z}\hat{%
\delta }_{zz^{\prime }},
\end{aligned}
\end{equation}%
which is satisfied by \eqref{inverse 0} above.
It follows that
\begin{equation}
\begin{aligned}
& \left\langle
\! \left\{ \phi_{\sigma }^{(0)}(\sigma
^{\prime \prime },z^{\prime \prime })_{\mathbf{3}},\phi_{%
\overline{z}}^{(0)}(\sigma ^{\prime \prime \prime },z^{\prime \prime \prime })_{%
\mathbf{4}}\right\} ^{-1},\left\{ \phi_{\overline{z}}^{(0)}(\sigma ^{\prime \prime
\prime },z^{\prime \prime \prime })_{\mathbf{4}},P_{\sigma}^{(0)}(\sigma ^{\prime
},z^{\prime })_{\mathbf{2}}\right\}\! \right \rangle _{(\sigma ^{\prime \prime
\prime },z^{\prime \prime \prime })\mathbf{4}}\\
&\quad =\frac{1}{2}C_{\mathbf{32}}%
^{(00)} \delta_{\sigma^{\prime \prime}\sigma^{\prime}}\hat{\delta }_{z''z^{\prime }}
\end{aligned}
\end{equation}
and 
\begin{equation}
\big\{A_{\sigma}^{(0)}(\sigma,z)_{\mathbf{1}},P_{\sigma}^{(0)}(\sigma^{\prime},z^{\prime})_{\mathbf{2}}\big\}^{\ast}=\frac{1}{2}\big\{A_{\sigma}^{(0)}(\sigma,z)_{\mathbf{1}},P_{\sigma}^{(0)}(\sigma^{\prime},z^{\prime})_{\mathbf{2}}\big\}.
\end{equation}

\textbf{Grade two sector.} 

In the same way as before, we need to compute
\begin{multline}
\! \! \! \! \! \! \big\{A_{\overline{z}}^{(2)}(\sigma,z)_{\mathbf{1}},P_{\overline{z}}^{(2)}(\sigma^{\prime},z^{\prime})_{\mathbf{2}}\big\}^{\ast}=\big\{A_{\overline{z}}^{(2)}(\sigma,z)_{\mathbf{1}},P_{\overline{z}}^{(2)}(\sigma^{\prime},z^{\prime})_{\mathbf{2}}\big\}\\
\! \! \! \! \! \! \! \! \! \! \! \! \! \! \! \! \! \! \! \! \! \! \! \! \! \! \! \! \! \! \! \! \! \! \! \! \! \! \! \! \! \! \! \! \! \! \! \! 
\! \! \! \! \! \! \! \! \! \! \! \! \! \! \! \! \! \! \! \! \! \! \! \! \! \! \! \! \! \! \! \! \! \! \! \! \! \! \! \! \! \! \! \! \! \! \! \! 
\! \! \! \! \! \! \! \! \! \! \! \! \! \! \! \! \! \! \! \! \! \! \! \! \! \! \! \! \! \!
\! \! \! \! \! \! \! \! \! \! \! \! \! \! \! \! \! \! \! \! \! \! \! \! \! \! \! \! \! \! \! \!
-\bigg\langle \! \left\{ A_{\overline{z}}^{(2)}(\sigma,z)_{\mathbf{1}},\phi_{\overline{z} }^{(2)}(\sigma
^{\prime \prime },z^{\prime \prime })_{\mathbf{3}}\right\} ,\\
\qquad \qquad \Big\langle
\! \left\{ \phi_{\overline{z} }^{(2)}(\sigma
^{\prime \prime },z^{\prime \prime })_{\mathbf{3}},\phi_{%
\sigma}^{(2)}(\sigma ^{\prime \prime \prime },z^{\prime \prime \prime })_{%
\mathbf{4}}\right\} ^{-1},\left\{ \phi_{\sigma}^{(2)}(\sigma ^{\prime \prime
\prime },z^{\prime \prime \prime })_{\mathbf{4}},P_{\overline{z}}^{(2)}(\sigma ^{\prime
},z^{\prime })_{\mathbf{2}}\right\}\! \Big\rangle _{(\sigma ^{\prime \prime
\prime },z^{\prime \prime \prime })\mathbf{4}}\bigg\rangle _{(\sigma
^{\prime \prime },z^{\prime \prime })\mathbf{3}}.
\end{multline}
From \eqref{primary}, the grade two constraint algebra is
\begin{equation}
\left\{ \phi _{\sigma }^{(2)}(\sigma ,z)_{\mathbf{1}},\phi _{_{\overline{z}%
}}^{(2)}(\sigma ^{\prime },z^{\prime })_{\mathbf{2}}\right\} =-\frac{1}{2}%
ic\varphi (z^{\prime })C_{\mathbf{12}}^{(22)}\delta _{\sigma \sigma
^{\prime }}\left( \frac{z}{z^{\prime }}%
\right) ^{2s}\hat{\delta }_{zz^{\prime }}.
\end{equation}%
Its inverse is defined by the covariant condition
\begin{equation}
\begin{aligned}
& \left\langle \left\{ \phi _{\sigma }^{(0)}(\sigma ,z)_{\mathbf{1}},\phi _{_{%
\overline{z}}}^{(0)}(\sigma ^{\prime \prime },z^{\prime \prime })_{\mathbf{3}%
}\right\} ,\left\{ \phi _{_{\overline{z}}}^{(0)}(\sigma ^{\prime \prime
},z^{\prime \prime })_{\mathbf{3}},\phi _{\sigma }^{(0)}(\sigma ^{\prime
},z^{\prime })_{\mathbf{2}}\right\} ^{-1}\right\rangle_{(\sigma^{\prime \prime},z^{\prime \prime})\mathbf{3}}\\
&\quad =C_{\mathbf{12}%
}^{(00)}\delta _{\sigma \sigma ^{\prime }}\left( \frac{z}{z^{\prime }}%
\right) ^{2s}\hat{\delta }_{zz^{\prime }}.
\end{aligned}
\end{equation}%
We find that%
\begin{equation}
\left\{ \phi _{_{\overline{z}}}^{(2)}(\sigma ,z)_{\mathbf{1}},\phi _{\sigma }^{(2)}(\sigma ^{\prime },z^{\prime })_{%
\mathbf{2}},\right\} ^{-1}=\frac{i}{2c}\varphi (z)^{-1}C_{\mathbf{12}%
}^{(22)}\delta _{\sigma \sigma ^{\prime }}\left( \frac{z}{z^{\prime }}%
\right) ^{2s}\hat{\delta }_{zz^{\prime }}.
\end{equation}%
From this follows
\begin{equation}
\begin{aligned}
& \left\langle
\! \left\{ \phi_{\overline{z} }^{(2)}(\sigma
^{\prime \prime },z^{\prime \prime })_{\mathbf{3}},\phi_{%
\sigma}^{(2)}(\sigma ^{\prime \prime \prime },z^{\prime \prime \prime })_{%
\mathbf{4}}\right\} ^{-1},\left\{ \phi_{\sigma}^{(2)}(\sigma ^{\prime \prime
\prime },z^{\prime \prime \prime })_{\mathbf{4}},P_{\overline{z}}^{(2)}(\sigma ^{\prime
},z^{\prime })_{\mathbf{2}}\right\}\! \right\rangle _{(\sigma ^{\prime \prime
\prime },z^{\prime \prime \prime })\mathbf{4}}\\
&\quad =\frac{1}{2}C_{\mathbf{32}}%
^{(22)} \delta_{\sigma^{\prime \prime}\sigma^{\prime}}\frac{z'}{z''}\hat{\delta }_{z''z^{\prime }}.
\end{aligned}
\end{equation}
Hence,
\begin{equation}
\big\{A_{\overline{z}}^{(2)}(\sigma,z)_{\mathbf{1}},P_{\overline{z}}^{(2)}(\sigma^{\prime},z^{\prime})_{\mathbf{2}}\big\}^{\ast}=\frac{1}{2}\big\{A_{\overline{z}}^{(2)}(\sigma,z)_{\mathbf{1}},P_{\overline{z}}^{(2)}(\sigma^{\prime},z^{\prime})_{\mathbf{2}}\big\}.
\end{equation}

Following the same logic, we need to compute
\begin{multline}
\! \! \! \! \! \! \big\{A_{\sigma}^{(2)}(\sigma,z)_{\mathbf{1}},P_{\sigma}^{(2)}(\sigma^{\prime},z^{\prime})_{\mathbf{2}}\big\}^{\ast}=\big\{A_{\sigma}^{(2)}(\sigma,z)_{\mathbf{1}},P_{\sigma}^{(2)}(\sigma^{\prime},z^{\prime})_{\mathbf{2}}\big\}\\
\! \! \! \! \! \! \! \! \! \! \! \! \! \! \! \! \! \! \! \! \! \! \! \! \! \! \! \! \! \! \! \! \! \! \! \! \! \! \! \! \! \! \! \! \! \! \! \! 
\! \! \! \! \! \! \! \! \! \! \! \! \! \! \! \! \! \! \! \! \! \! \! \! \! \! \! \! \! \! \! \! \! \! \! \! \! \! \! \! \! \! \! \! \! \! \! \! 
\! \! \! \! \! \! \! \! \! \! \! \! \! \! \! \! \! \! \! \! \! \! \! \! \! \! \! \! \! \!
\! \! \! \! \! \! \! \! \! \! \! \! \! \! \! \! \! \! \! \! \! \! \! \! \! \! \! \! \! \! \! \!
-\bigg\langle \! \left\{ A_{\sigma}^{(2)}(\sigma,z)_{\mathbf{1}},\phi_{\sigma}^{(2)}(\sigma
^{\prime \prime },z^{\prime \prime })_{\mathbf{3}}\right\} ,\\
\qquad \qquad \Big\langle
\! \left\{ \phi_{\sigma }^{(2)}(\sigma
^{\prime \prime },z^{\prime \prime })_{\mathbf{3}},\phi_{%
\overline{z}}^{(2)}(\sigma ^{\prime \prime \prime },z^{\prime \prime \prime })_{%
\mathbf{4}}\right\} ^{-1},\left\{ \phi_{\overline{z}}^{(2)}(\sigma ^{\prime \prime
\prime },z^{\prime \prime \prime })_{\mathbf{4}},P_{\sigma}^{(2)}(\sigma ^{\prime
},z^{\prime })_{\mathbf{2}}\right\}\! \Big\rangle _{(\sigma ^{\prime \prime
\prime },z^{\prime \prime \prime })\mathbf{4}}\bigg\rangle _{(\sigma
^{\prime \prime },z^{\prime \prime })\mathbf{3}}.
\end{multline}
This time, the inverse is defined by the covariant condition
\begin{equation}
\begin{aligned}
& \left\langle \left\{ \phi _{_{\overline{z}}}^{(2)}(\sigma ,z)_{\mathbf{1}%
},\phi _{_{\sigma }}^{(2)}(\sigma ^{\prime \prime },z^{\prime \prime })_{%
\mathbf{3}}\right\} ,\left\{ \phi _{\sigma }^{(2)}(\sigma ^{\prime \prime
},z^{\prime \prime })_{\mathbf{3}},\phi _{_{\overline{z}}}^{(2)}(\sigma
^{\prime },z^{\prime })_{\mathbf{2}},\right\} ^{-1}\right\rangle _{(\sigma
^{\prime \prime },z^{\prime \prime })\mathbf{3}}\\
&\quad =C_{\mathbf{12}%
}^{(22)}\delta _{\sigma \sigma ^{\prime }}\frac{z}{z'}\hat{%
\delta }_{zz^{\prime }}.
\end{aligned}
\end{equation}%
It follows that
\begin{equation}
\begin{aligned}
& \left\langle
\! \left\{ \phi_{\sigma }^{(2)}(\sigma
^{\prime \prime },z^{\prime \prime })_{\mathbf{3}},\phi_{%
\overline{z}}^{(2)}(\sigma ^{\prime \prime \prime },z^{\prime \prime \prime })_{%
\mathbf{4}}\right\} ^{-1},\left\{ \phi_{\overline{z}}^{(2)}(\sigma ^{\prime \prime
\prime },z^{\prime \prime \prime })_{\mathbf{4}},P_{\sigma}^{(2)}(\sigma ^{\prime
},z^{\prime })_{\mathbf{2}}\right\}\! \right\rangle _{(\sigma ^{\prime \prime
\prime },z^{\prime \prime \prime })\mathbf{4}}\\
&\quad =\frac{1}{2}C_{\mathbf{32}}%
^{(22)} \delta_{\sigma^{\prime \prime}\sigma^{\prime}}\left( \frac{z''}{z^{\prime }}%
\right) ^{2s}\hat{\delta }_{z''z^{\prime }}
\end{aligned}
\end{equation}
and 
\begin{equation}
\big\{A_{\sigma}^{(2)}(\sigma,z)_{\mathbf{1}},P_{\sigma}^{(2)}(\sigma^{\prime},z^{\prime})_{\mathbf{2}}\big\}^{\ast}=\frac{1}{2}\big\{A_{\sigma}^{(2)}(\sigma,z)_{\mathbf{1}},P_{\sigma}^{(2)}(\sigma^{\prime},z^{\prime})_{\mathbf{2}}\big\}.
\end{equation}

Finally, the Dirac bracket is one-half the original Poisson bracket and now we impose the second class constrains $\phi_{\sigma}=\phi_{\overline{z}}=0$ strongly. The resulting bracket inherits all the equivariance properties of the components fields and it is given by
\begin{equation}
\begin{aligned}
\left\{ A_{\overline{z}}(\sigma ,z)_{\mathbf{1}},A_{\sigma }(\sigma ^{\prime
},z^{\prime })_{\mathbf{2}}\right\} ^{\ast } &=\frac{i}{8c}\varphi
(z^{\prime })^{-1}\left( \frac{z}{z^{\prime }}C_{\mathbf{12}}^{(00)}+\frac{z^{\prime }}{z}C_{%
\mathbf{12}}^{(22)}\right) \hat{\delta }%
_{zz^{\prime }}\delta _{\sigma \sigma ^{\prime }}, \\
\left\{ A_{\sigma }(\sigma ,z)_{\mathbf{1}},A_{\overline{z}}(\sigma ^{\prime
},z^{\prime })_{\mathbf{2}}\right\} ^{\ast } &=-\frac{i}{8c}\varphi
(z)^{-1}\left( \frac{z^{\prime }}{z}C_{\mathbf{12}}^{(00)}+\frac{z}{z^{\prime }}C_{\mathbf{12}%
}^{(22)}\right) \hat{\delta }_{zz^{\prime }}\delta
_{\sigma \sigma ^{\prime }}.
\end{aligned}
\end{equation}%
In this presentation, these expressions do not tell much. However, once
we work out the delta distributions, we obtain a remarkable result 
\begin{equation}
\begin{aligned}
\left\{ A_{\overline{z}}(\sigma ,z)_{\mathbf{1}},A_{\sigma }(\sigma ^{\prime
},z^{\prime })_{\mathbf{2}}\right\} ^{\ast } &=\frac{1}{4}\partial _{%
\overline{z}}R_{\mathbf{12}}(z,z^{\prime })\delta _{\sigma \sigma ^{\prime
}}, \\
\left\{ A_{\sigma }(\sigma ,z)_{\mathbf{1}},A_{\overline{z}}(\sigma ^{\prime
},z^{\prime })_{\mathbf{2}}\right\} ^{\ast } &=-\frac{1}{4}\partial _{%
\overline{z}^{\prime }}R_{\mathbf{12}}^{\ast }(z,z^{\prime })\delta _{\sigma
\sigma ^{\prime }},
\end{aligned}
\end{equation}%
where%
\begin{equation}
\begin{aligned}
R_{\mathbf{12}}(z,z^{\prime }) &=-\frac{2z^{\prime 4}}{%
z^{4}-z^{\prime 4}}\phi (z^{\prime })^{-1}\Omega _{\mathbf{1}}^{-1}(z)\Omega
_{\mathbf{2}}(z^{\prime })C_{\mathbf{12}}, \\
R_{\mathbf{12}}^{\ast }(z,z^{\prime }) &=-\frac{2z^{4}}{%
z^{\prime 4}-z^{4}}\phi (z)^{-1}\Omega _{\mathbf{1}}(z)\Omega _{\mathbf{2}%
}^{-1}(z^{\prime })C_{\mathbf{12}}, \label{R-matrix def}
\end{aligned}
\end{equation}%
are the R-matrix and its adjoint and $\phi(z) =z\varphi(z) $ is the $\mathbb{Z}_{4}$-invariant twist function. Notice that  $R_{\mathbf{12}}^{\ast }(z,z^{\prime
})=R_{\mathbf{21}}(z^{\prime },z)$ in agreement with the properties of the bracket. Here is where the R-matrix of the symmetric space lambda model first appear. Above, we have set
\begin{equation}
c=1/2\pi.
\end{equation}

Summarizing, the complete Dirac bracket, at this stage, is given by
\begin{equation}
\left\{ f,g\right\} ^{\ast }=\frac{1}{4}\left\langle 
\begin{array}{c}
\frac{\delta f}{\delta A_{\tau }(\sigma ,z)},\Phi ^{\alpha }\left( \frac{%
\delta g}{\delta P_{\tau }(\sigma ,i^{\alpha }z)}\right) -\frac{\delta f}{%
\delta P_{\tau }(\sigma ,z)},\Phi ^{\alpha }\left( \frac{\delta g}{\delta
A_{\tau }(\sigma ,i^{\alpha }z)}\right)  \\ 
+\frac{i}{2c}\varphi (z)^{-1}\left( \frac{\delta f}{\delta A_{\overline{z}%
}(\sigma ,z)},\Phi ^{\alpha }\left( \frac{\delta g}{\delta A_{\sigma
}(\sigma ,i^{\alpha }z)}\right) -\Phi ^{\alpha }\left( \frac{\delta f}{%
\delta A_{\sigma }(\sigma ,i^{\alpha }z)}\right) , \frac{\delta g}{%
\delta A_{\overline{z}}(\sigma ,z)} \right) 
\end{array}%
\right\rangle _{(\sigma ,z)}.\label{Dirac 1}
\end{equation}%
The remaining two constraints are, one primary and one secondary,
\begin{equation}
P_{\tau }\approx 0,\qquad G_{0}(\eta )\approx 0, \label{P and G}
\end{equation}%
while the total Hamiltonian is now%
\begin{equation}
h_{T}=h+\left\langle u_{\tau },P_{\tau }\right\rangle _{(\sigma ,z)}.
\end{equation}

\subsubsection{Emergence of the exchange algebra}

Here we continue with the process of gauge fixing the CS theory. The main result at this stage is the emergence of the coset lambda model exchange algebra, which takes the form of the Maillet bracket. The non-ultralocality of the integrable field theory shows for the first time. 

Both constraints \eqref{P and G} are actually first class, with $G_{0}(\eta)$ generating gauge transformations. Indeed, we have that 
\begin{equation}
\big\{G_{0}(\eta),G_{0}(\overline{\eta}) \big\}^{\ast}\approx 0,\qquad \big\{G_{0}(\eta),A_{k} \big\}^{\ast}=-D_{k}\eta, \label{gauge trans}
\end{equation}
for $k=\sigma,\overline{z}$. The compatibility of the gauge group action and the conditions \eqref{comp-trans} is direct, once we take into account \eqref{gauge par equiv} and the fact that $\Phi$ is an automorphism. The gauge group action preserves the equivariance properties of the gauge field components, i.e. taking $z\rightarrow iz$ in the second expression of \eqref{gauge trans}, gives
\begin{equation}
\left\{ G_{0}(\eta ),A_{\sigma }(\tau ,\sigma ,iz)\right\}^{\ast}  =-\Phi 
D_{\sigma }\eta  (\tau ,\sigma ,z), \qquad \left\{ G_{0}(\eta ),A_{\overline{z}}(\tau ,\sigma ,iz)\right\}^{\ast}  =-i\Phi
 D_{_{\overline{z}}}\eta  (\tau ,\sigma ,z).
\end{equation}
Conversely, the bracket between $G_{0}(\eta)$ and the expressions in \eqref{comp-trans}, gives the same result as above. Both group actions commute.

The gauge fixing condition for $G_{0}(\eta)$ is chosen, recall the discussion around \eqref{condition ove Lax}, to be
\begin{equation}
A_{\overline{z}}\approx 0. \label{gauge-fixing A}
\end{equation} 
It is preserved under gauge symmetries and also in time provided we have, respectively,
\begin{equation}
\left\{G_{0}(\eta),A_{\overline{z}}(z)\right\}\approx -\partial_{\overline{z}}\eta(z)\approx 0,\qquad \varphi(z)\left\{h,A_{\overline{z}}(z)   \right\}\approx -\varphi(z)\partial_{\overline{z}}A_{\tau}(z)\approx 0.
\end{equation}
The first condition means that $\eta$ is restricted to be holomorphic in order for \eqref{gauge-fixing A} to remain zero as chosen, the second condition is equivalent to \eqref{condition ove Lax} for $\mu=\tau$. The latter being satisfied when $A_{\tau}\approx \mathscr{L}_{\tau}$. Notice that, we have to multiply with the twist function in order to reproduce the Euler-Lagrange eom \eqref{full eom} from the Hamiltonian formalism, e.g. 
\begin{equation}
\partial_{\tau}A_{\sigma}\equiv -\left\{h,A_{\sigma} \right\}= D_{\sigma}A_{\tau},\qquad  \partial_{\tau}A_{\overline{z}}\equiv - \left\{h,A_{\overline{z}} \right\}= D_{\overline{z}}A_{\tau},
\end{equation}
are equivalent to the first and second eom (with $\mu=\tau$) of \eqref{full eom} after multiplication by $\varphi$.

Now, we proceed to compute the Dirac bracket for the pair of second class constraints
\begin{equation}
\gamma=2ic\varphi F_{\overline{z}\sigma}\approx 0,\qquad A_{\overline{z}}\approx 0.  \label{pair}
\end{equation} 
The only non-trivial Dirac bracket is given by
\begin{multline}
\! \! \! \! \! \! \big\{A_{\sigma}(\sigma,z)_{\mathbf{1}},A_{\sigma}(\sigma^{\prime},z^{\prime})_{\mathbf{2}}\big\}^{\star}=\big\{A_{\sigma}(\sigma,z)_{\mathbf{1}},A_{\sigma}(\sigma^{\prime},z^{\prime})_{\mathbf{2}}\big\}^{\ast}\\
\! \! \! \! \! \! \! \! \! \! \! \! \! \! \! \! \! \! \! \! \! \! \! \! \! \! \! \! \! \! \! \! \! \! \! \! \! \! \! \! \! \! \! \! \! \! \! \! 
\! \! \! \! \! \! \! \! \! \! \! \! \! \! \! \! \! \! \! \! \! \! \! \! \! \! \! \! \! \! \! \! \! \! \! \! \! \! \! \! \! \! \! \! \! \! \! \! 
\! \! \! \! \! \! \! \! \! \! \! \! \! \! \! \! \! \! \! \! \! \! \! \! \! \! \! \! \! \!
\! \! \! \! \! \! \! \! \! \! \! \! \! \! \! \! \! \! \! \! \! \! \! \! \! \! \! \! \! \! \! \! \! \! \! \! \! \! \! \! \!  
-\bigg\langle \! \left\{ A_{\sigma }(\sigma
,z)_{\mathbf{1}},\gamma (\sigma^{\prime \prime},z^{\prime \prime})_{\mathbf{3}}\right\}^{\ast} , \\
\qquad \qquad \qquad \quad \; \Big\langle \!
\left\{ \gamma (\sigma ^{\prime \prime },z^{\prime \prime })_{\mathbf{3}},A_{%
\overline{z}}(\sigma ^{\prime \prime \prime },z^{\prime \prime \prime })_{%
\mathbf{4}}\right\} ^{\ast-1},\left\{ A_{\overline{z}}(\sigma ^{\prime \prime
\prime },z^{\prime \prime \prime })_{\mathbf{4}},A_{\sigma }(\sigma ^{\prime
},z^{\prime })_{\mathbf{2}}\right\}^{\ast} \! \Big \rangle _{(\sigma ^{\prime \prime
\prime },z^{\prime \prime \prime })\mathbf{4}}\bigg\rangle _{(\sigma
^{\prime \prime },z^{\prime \prime })\mathbf{3}} \\
 \! \! \! \! \! \! \! \! \! \! \! \! \! \! \! \! \! \! \! \!  \! \! \! \! \! \! \! \! \! \! \! \! \! \! \! \! \! \! \! \!
 \! \! \! \! \! \! \! \! \! \! \! \! \! \! \! \! \! \! \! \!  \! \! \! \! \! \! \! \! \! \! \! \! \! \! \! \! \! \! \! \!
 \! \! \! \! \! \! \! \! \! \! \! \! \! \! \! \! \! \! \! \!  \! \! \! \! \! \! \! \! \! \! \! \! \! \! \! \! \! \! \! \! \!
 \! \! \! \! \! \! \! \! \! \! \! \! \! \! \! \! \! \! \! \!  \! \! \! \! \!  \! \! \! \!  \! \! \! \! \! \! \!  \! \! \! \!  \! \! \!
 -\bigg\langle \! \left\{ A_{\sigma }(\sigma ,z)_{\mathbf{1}},A_{\overline{z}%
}(\sigma ^{\prime \prime },z^{\prime \prime })_{\mathbf{3}}\right\}^{\ast}
, \\
\Big\langle \! \left\{ A_{\overline{z}}(\sigma ^{\prime \prime },z^{\prime
\prime })_{\mathbf{3}},\gamma (\sigma ^{\prime \prime \prime },z^{\prime
\prime \prime })_{\mathbf{4}}\right\} ^{\ast-1},\left\{ \gamma (\sigma ^{\prime
\prime \prime },z^{\prime \prime \prime })_{\mathbf{4}},A_{\sigma }(\sigma
^{\prime },z^{\prime })_{\mathbf{2}}\right\}^{\ast}\! \Big\rangle _{(\sigma
^{\prime \prime \prime },z^{\prime \prime \prime })\mathbf{4}}\bigg\rangle
_{(\sigma ^{\prime \prime },z^{\prime \prime })\mathbf{3}}.
\end{multline}
After exchanging the orders of $(\sigma
^{\prime \prime },z^{\prime \prime })\mathbf{3}$ and $(\sigma ^{\prime
\prime \prime },z^{\prime \prime \prime })\mathbf{4}$ in the last term on the right hand side and remembering that the first contribution to the bracket is zero, we have 
\begin{multline}
\! \! \! \! \! \! \big\{A_{\sigma}(\sigma,z)_{\mathbf{1}},A_{\sigma}(\sigma^{\prime},z^{\prime})_{\mathbf{2}}\big\}^{\star}=\\
\! \! \! \! \! \! \! \! \! \! \! \! \! \! \! \! \! \! \! \! \! \! \! \! \! \! \! \! \! \! \! \! \! \! \! \! \! \! \! \! \! \! \! \! \! \! \! \! 
\! \! \! \! \! \! \! \! \! \! \! \! \! \! \! \! \! \! \! \! \! \! \! \! \! \! \! \! \! \! \! \! \! \! \! \! \! \! \! \! \! \! \! \! \! \! \! \! 
\! \! \! \! \! \! \! \! \! \! \! \! \! \! \! \! \! \! \! \! \! \! \! \! \! \! \! \! \! \!
\! \! \! \! \! \! \! \! \! \! \! \! \! \! \! \! \! \! \! \! \! \! \! \! \! \! \! \! \! \! \! \! \! \! \! \! \! \! \! \! \! 
-\bigg\langle \! \left\{ \gamma (\sigma ,z)_{\mathbf{1}},A_{\sigma }(\sigma
^{\prime \prime },z^{\prime \prime })_{\mathbf{3}}\right\} ,\\
\qquad \qquad \qquad \quad \; \Big\langle
\! \left\{ \gamma (\sigma ^{\prime \prime },z^{\prime \prime })_{\mathbf{3}},A_{%
\overline{z}}(\sigma ^{\prime \prime \prime },z^{\prime \prime \prime })_{%
\mathbf{4}}\right\} ^{-1},\left\{ A_{\overline{z}}(\sigma ^{\prime \prime
\prime },z^{\prime \prime \prime })_{\mathbf{4}},A_{\sigma }(\sigma ^{\prime
},z^{\prime })_{\mathbf{2}}\right\}\! \Big\rangle _{(\sigma ^{\prime \prime
\prime },z^{\prime \prime \prime })\mathbf{4}}\bigg\rangle _{(\sigma
^{\prime \prime },z^{\prime \prime })\mathbf{3}} \\
\! \! \! \! \! \! \! \! \! \! \! \! \! \! \!  \! \! \! \! \! \! \! \! \! \! \! \! \! \! \! \! \! \! \! \! \! \! \! \! \! \! \! \! \! \! \! \! \! \! \! \! \! \! \! \! \! \! \! \! \! \!  -\bigg\langle  \Big\langle \! \left\{ A_{\sigma }(\sigma ,z)_{\mathbf{1}},A_{%
\overline{z}}(\sigma ^{\prime \prime },z^{\prime \prime })_{\mathbf{3}%
}\right\} ,\left\{ A_{\overline{z}}(\sigma ^{\prime \prime },z^{\prime
\prime })_{\mathbf{3}},\gamma (\sigma ^{\prime \prime \prime },z^{\prime
\prime \prime })_{\mathbf{4}}\right\} ^{-1} \! \Big\rangle _{(\sigma ^{\prime
\prime },z^{\prime \prime })\mathbf{3}}, \\ 
\quad \qquad \qquad \qquad  \left\{ \gamma (\sigma ^{\prime \prime \prime },z^{\prime \prime \prime })_{\mathbf{4}},A_{\sigma }(\sigma
^{\prime },z^{\prime })_{\mathbf{2}}\right\}\! \bigg\rangle _{(\sigma
^{\prime \prime \prime },z^{\prime \prime \prime })\mathbf{4}}.
\end{multline}

The key relation now to be considered is the covariant bracket
\begin{equation}
\left\{ \gamma (\sigma ,z)_{\mathbf{1}},A_{\overline{z}}(\sigma ^{\prime
},z^{\prime })\right\}^{\ast} \approx -\frac{1}{4}\left\{ 
C_{\mathbf{12}%
}^{(00)}+\left(\frac{z'}{z}\right)^{2}  C_{\mathbf{%
12}}^{(22)}\right\}\partial _{\overline{z}'}\hat{\delta }_{zz^{\prime }} \delta _{\sigma \sigma ^{\prime }}. \label{I}
\end{equation}%
The calculation of \eqref{I} depends on using in \eqref{Dirac 1}, the term
\begin{equation}
\Phi _{\mathbf{3}}^{\alpha }\left( \frac{\delta \gamma (\sigma ,z)_{\mathbf{1%
}}}{\delta A_{\sigma }(\sigma ^{\prime \prime },i^{\alpha }z^{\prime \prime
})_{\mathbf{3}}}\right) \approx 2ic\varphi (z)\partial _{\overline{z}}\left\{ C_{\mathbf{13}}^{(00)}\hat{\delta }%
_{zz^{\prime \prime }}+\left( \frac{z}{z^{\prime \prime }}\right) ^{2s}C_{\mathbf{13}}^{(22)}%
\hat{\delta }_{zz^{\prime \prime }}\right\}\delta _{\sigma \sigma ^{\prime \prime
}} 
\end{equation}%
and the distribution equation\footnote{To show this, act on both sides with $\int dz\wedge d\overline{z}f(z)$, with $f(z)$ a test function.}%
\begin{equation}
\partial _{\overline{z}}\delta (z-pz^{\prime })=-(1/\overline{p})\partial _{%
\overline{z}^{\prime }}\delta (z-pz^{\prime }),
\end{equation}%
with $p\in \mathbb{C}$, in order to obtain%
\begin{equation}
\left\{ \gamma (\sigma ,z)_{\mathbf{1}},A_{\overline{z}}(\sigma ^{\prime
},z^{\prime })_{\mathbf{2}}\right\} ^{\ast }\approx -\frac{1}{4}\left\{ C_{\mathbf{12}}^{(00)}+\left( \frac{z^{\prime }}{z}%
\right) ^{2}C_{\mathbf{12}}^{(22)}\right\} \phi
(z^{\prime })^{-1}\partial _{\overline{z}^{\prime }}\left(%
\phi (z)\hat{\delta }_{zz^{\prime }}\right)\delta _{\sigma \sigma ^{\prime }}.
\end{equation}

As done above with our first Dirac bracket, we compute the inverse of this quantity by sectors. 

\textbf{Grade zero sector}.

The inverse is defined by the covariant condition%
\begin{equation}
\begin{aligned}
& \left\langle \left\{ A_{\overline{z}}^{(0)}(\sigma ,z)_{\mathbf{1}},\gamma
^{(0)}(\sigma ^{\prime \prime },z^{\prime \prime })_{\mathbf{3}}\right\}^{\ast}
,\left\{ \gamma ^{(0)}(\sigma ^{\prime \prime },z^{\prime \prime })_{\mathbf{%
3}},A_{\overline{z}}^{(0)}(\sigma ^{\prime },z^{\prime })_{\mathbf{2}%
}\right\} ^{*-1}\right\rangle _{(\sigma ^{\prime \prime },z^{\prime \prime
})_{\mathbf{3}}}\\
&\quad =C_{\mathbf{12}}^{(00)}\delta _{\sigma \sigma ^{\prime }}%
\frac{z}{z^{\prime }}\hat{\delta }_{zz^{\prime }}.
\end{aligned}
\end{equation}%
We find that%
\begin{equation}
\left\{ \gamma ^{(0)}(\sigma ,z)_{\mathbf{1}},A_{\overline{z}}^{(0)}(\sigma
^{\prime },z^{\prime })_{\mathbf{2}}\right\} ^{*-1}=C_{\mathbf{12}%
}^{(00)}\delta _{\sigma \sigma ^{\prime }}f(z,z^{\prime }),
\end{equation}%
where%
\begin{equation}
f(z,z^{\prime })=-\frac{1}{2\pi i}\dsum\limits_{\alpha =0}^{3}\frac{i^{\alpha
}}{z-i^{\alpha }z^{\prime }},
\end{equation}%
satisfy the following relations%
\begin{equation}
f(iz,z^{\prime })=f(z,z^{\prime }),\qquad f(z,iz^{\prime
})=-if(z,z^{\prime }),\qquad \partial _{\overline{z}}f(z,z^{\prime })=%
\frac{z}{z^{\prime }}\hat{\delta }_{zz^{\prime }},\qquad \partial _{%
\overline{z}^{\prime }}f(z,z^{\prime })=-\hat{\delta }_{zz^{\prime }}.
\end{equation}
Then, we get
\begin{equation}
\begin{aligned}
& \left\langle \left\{ \gamma ^{(0)}(\sigma ^{\prime \prime },z^{\prime \prime
})_{\mathbf{3}},A_{\overline{z}}^{(0)}(\sigma ^{\prime \prime \prime
},z^{\prime \prime \prime })_{\mathbf{4}}\right\} ^{*-1},\left\{ A_{\overline{%
z}}^{(0)}(\sigma ^{\prime \prime \prime },z^{\prime \prime \prime })_{%
\mathbf{4}},A_{\sigma }^{(0)}(\sigma ^{\prime },z^{\prime })_{\mathbf{2}%
}\right\}^{\ast} \right\rangle _{(\sigma ^{\prime \prime \prime },z^{\prime \prime
\prime })_{\mathbf{4}}}\\
&\quad =R_{\mathbf{32}}^{(00)}(z^{\prime \prime
},z^{\prime })\delta _{\sigma ^{\prime \prime }\sigma ^{\prime }}
\end{aligned}
\end{equation}%
and also
\begin{equation}
\begin{aligned}
&\left\langle \left\{ A_{\sigma }^{(0)}(\sigma ,z)_{\mathbf{1}},A_{\overline{z%
}}^{(0)}(\sigma ^{\prime \prime },z^{\prime \prime })_{\mathbf{3}}\right\}^{\ast}
,\left\{ A_{\overline{z}}^{(0)}(\sigma ^{\prime \prime },z^{\prime \prime
})_{\mathbf{3}},\gamma ^{(0)}(\sigma ^{\prime \prime \prime },z^{\prime
\prime \prime })_{\mathbf{4}}\right\} ^{*-1}\right\rangle _{(\sigma ^{\prime
\prime },z^{\prime \prime })_{\mathbf{3}}}\\
&\quad =R_{\mathbf{14}}^{\ast
(00)}(z,z^{\prime \prime \prime })\delta _{\sigma \sigma ^{\prime \prime
\prime }}.
\end{aligned}
\end{equation}

\textbf{Grade two sector}.

The inverse is now defined by the covariant condition%
\begin{equation}
\begin{aligned}
& \left\langle \left\{ A_{\overline{z}}^{(2)}(\sigma ,z)_{\mathbf{1}},\gamma
^{(2)}(\sigma ^{\prime \prime },z^{\prime \prime })_{\mathbf{3}}\right\}^{\ast}
,\left\{ \gamma ^{(2)}(\sigma ^{\prime \prime },z^{\prime \prime })_{\mathbf{%
3}},A_{\overline{z}}^{(2)}(\sigma ^{\prime },z^{\prime })_{\mathbf{2}%
}\right\} ^{*-1}\right\rangle _{(\sigma ^{\prime \prime },z^{\prime \prime
})_{\mathbf{3}}}\\
&\quad =C_{\mathbf{12}}^{(00)}\delta _{\sigma \sigma ^{\prime }}%
\frac{z^{\prime }}{z}\hat{\delta }_{zz^{\prime }}.
\end{aligned}
\end{equation}%
We find%
\begin{equation}
\left\{ \gamma ^{(2)}(\sigma ,z)_{\mathbf{1}},A_{\overline{z}}^{(2)}(\sigma
^{\prime },z^{\prime })_{\mathbf{2}}\right\} ^{*-1}=C_{\mathbf{12}%
}^{(22)}\delta _{\sigma \sigma ^{\prime }}h(z,z^{\prime }),
\end{equation}%
where%
\begin{equation}
h(z,z^{\prime })=-\frac{1}{2\pi i}\dsum\limits_{\alpha =0}^{3}\frac{%
i^{-\alpha }}{z-i^{\alpha }z^{\prime }},
\end{equation}%
satisfy the relations%
\begin{equation}
h(iz,z^{\prime })=-h(z,z^{\prime }),\quad h(z,iz^{\prime
})=if(z,iz^{\prime }),\quad \partial _{\overline{z}}h(z,z^{\prime })=%
\frac{z^{\prime }}{z}\hat{\delta }_{zz^{\prime }},\quad \partial _{%
\overline{z}^{\prime }}h(z,z^{\prime })=-\left( \frac{z}{z^{\prime }}\right)
^{2s}\hat{\delta }_{zz^{\prime }}.
\end{equation}
Then, we have that%
\begin{equation}
\begin{aligned}
&\left\langle \left\{ \gamma ^{(2)}(\sigma ^{\prime \prime },z^{\prime \prime
})_{\mathbf{3}},A_{\overline{z}}^{(2)}(\sigma ^{\prime \prime \prime
},z^{\prime \prime \prime })_{\mathbf{4}}\right\} ^{*-1},\left\{ A_{\overline{%
z}}^{(2)}(\sigma ^{\prime \prime \prime },z^{\prime \prime \prime })_{%
\mathbf{4}},A_{\sigma }^{(2)}(\sigma ^{\prime },z^{\prime })_{\mathbf{2}%
}\right\}^{\ast} \right\rangle _{(\sigma ^{\prime \prime \prime },z^{\prime \prime
\prime })_{\mathbf{4}}}\\
&\quad =R_{\mathbf{32}}^{(22)}(z^{\prime \prime
},z^{\prime })\delta _{\sigma ^{\prime \prime }\sigma ^{\prime }}
\end{aligned}
\end{equation}%
and also%
\begin{equation}
\begin{aligned}
&\left\langle \left\{ A_{\sigma }^{(2)}(\sigma ,z)_{\mathbf{1}},A_{\overline{z%
}}^{(2)}(\sigma ^{\prime \prime },z^{\prime \prime })_{\mathbf{3}}\right\}^{\ast}
,\left\{ A_{\overline{z}}^{(2)}(\sigma ^{\prime \prime },z^{\prime \prime
})_{\mathbf{3}},\gamma ^{(2)}(\sigma ^{\prime \prime \prime },z^{\prime
\prime \prime })_{\mathbf{4}}\right\} ^{*-1}\right\rangle _{(\sigma ^{\prime
\prime },z^{\prime \prime })_{\mathbf{3}}}\\
&\quad =R_{\mathbf{14}}^{\ast
(22)}(z,z^{\prime \prime \prime })\delta _{\sigma \sigma ^{\prime \prime
\prime }}.
\end{aligned}
\end{equation}

Altogether gives the interesting result
\begin{equation}
\begin{aligned}
\left\langle \left\{ \gamma (\sigma ,z)_{\mathbf{1}},A_{\overline{z}}(\sigma
^{\prime \prime },z^{\prime \prime })_{\mathbf{3}}\right\} ^{*-1},\left\{ A_{%
\overline{z}}(\sigma ^{\prime \prime },z^{\prime \prime })_{\mathbf{3}%
},A_{\sigma }(\sigma ^{\prime },z^{\prime })_{\mathbf{2}}\right\}^{\ast}
\right\rangle _{(\sigma ^{\prime \prime },z^{\prime \prime })_{\mathbf{3}}}
&=R_{\mathbf{12}}(z,z^{\prime })\delta _{\sigma \sigma ^{\prime
}}, \\
\left\langle \left\{ A_{\sigma }(\sigma ,z)_{\mathbf{1}},A_{\overline{z}%
}(\sigma ^{\prime \prime },z^{\prime \prime })_{\mathbf{3}}\right\}^{\ast}
,\left\{ A_{\overline{z}}(\sigma ^{\prime \prime },z^{\prime \prime })_{%
\mathbf{3}},\gamma (\sigma ^{\prime },z^{\prime })_{\mathbf{2}}\right\}
^{*-1}\right\rangle _{(\sigma ^{\prime \prime },z^{\prime \prime })_{\mathbf{3%
}}} &=R_{\mathbf{12}}^{\ast }(z,z^{\prime })\delta _{\sigma
\sigma ^{\prime }}.
\end{aligned}
\end{equation}%
Inserting these expressions into the Dirac bracket above, we get that 
\begin{equation}
\begin{aligned}
\left\{ A_{\sigma }(\sigma ,z)_{\mathbf{1}},A_{\sigma }(\sigma ^{\prime
},z^{\prime })_{\mathbf{2}}\right\} ^{\star} =\; & -\left\langle
\left\{\gamma (\sigma ,z)_{\mathbf{1}},A_{\sigma }(\sigma ^{\prime \prime
},z^{\prime \prime })_{\mathbf{3}}\right\}^{\ast} ,R_{\mathbf{32}}(z^{\prime \prime
},z^{\prime })\delta _{\sigma ^{\prime \prime }\sigma ^{\prime
}}\right\rangle _{(\sigma ^{\prime \prime },z^{\prime \prime })\mathbf{3}} \\
&\quad -\left\langle R_{\mathbf{13}}^{\ast }(z,z^{\prime \prime
})\delta _{\sigma \sigma ^{\prime \prime }},\left\{ \gamma (\sigma ^{\prime
\prime },z^{\prime \prime })_{\mathbf{3}},A_{\sigma }(\sigma ^{\prime
},z^{\prime })_{\mathbf{2}}\right\}^{\ast} \right\rangle _{(\sigma ^{\prime \prime
},z^{\prime \prime })\mathbf{3}}.
\end{aligned}
\end{equation}

Now, the ingredient required to simplify the bracket right above is\footnote{We have defined $\delta_{\sigma \sigma'}^{\prime}=\partial_{\sigma}\delta(\sigma-\sigma')$. This CS gauge symmetry transformation is responsible for the lambda model exchange algebra non-ultralocality term.}
\begin{equation}
\begin{aligned}
&\left\{ \gamma (\sigma ,z)_{\mathbf{1}},A_{\sigma }(\sigma ^{\prime
},z^{\prime })_{\mathbf{2}}\right\}^{\ast}\\
& =\frac{1}{4}\left\{ 
\begin{array}{c}
C_{\mathbf{12}}^{(00)}\delta _{\sigma \sigma ^{\prime }}^{\prime }+\left[ C_{%
\mathbf{12}}^{(00)},A_{\sigma }^{(0)}(\sigma ,z)_{\mathbf{2}}\right] \delta
_{\sigma \sigma ^{\prime }}+\left[ C_{\mathbf{12}}^{(22)},A_{\sigma
}^{(2)}(\sigma ,z)_{\mathbf{2}}\right] \delta _{\sigma \sigma ^{\prime }} \\ 
+(\frac{z}{z^{\prime }})^{2s}\left(C_{%
\mathbf{12}}^{(22)}\delta _{\sigma \sigma ^{\prime }}^{\prime }+\left[ C_{\mathbf{12}}^{(00)},A_{\sigma }^{(2)}(\sigma
,z)_{\mathbf{2}}\right] \delta _{\sigma \sigma ^{\prime }}+\left[ C_{\mathbf{%
12}}^{(22)},A_{\sigma }^{(0)}(\sigma ,z)_{\mathbf{2}}\right] \delta _{\sigma
\sigma ^{\prime }}\right)%
\end{array}%
\right\} \hat{\delta }_{zz^{\prime }}.
\end{aligned}
\end{equation}%
From this expression and after computing the bracket grade by grade, we obtain
\begin{equation*}
\begin{aligned}
\left\{ A_{\sigma }^{(0)}(\sigma ,z)_{\mathbf{1}},A_{\sigma }^{(0)}(\sigma
^{\prime },z^{\prime })_{\mathbf{2}}\right\} ^{\star} =-R_{%
\mathbf{12}}^{(00)}(z,&z^{\prime })\delta _{\sigma \sigma ^{\prime }}^{\prime
}+\left[ R_{\mathbf{12}}^{(00)}(z,z^{\prime }),A_{\sigma }^{(0)}(\sigma ,z)_{%
\mathbf{1}}\right] \delta _{\sigma \sigma ^{\prime }} \\
& -R_{\mathbf{12}}^{\ast (00)}(z,z^{\prime })\delta _{\sigma \sigma ^{\prime
}}^{\prime }-\left[ R_{\mathbf{12}}^{\ast (00)}(z,z^{\prime }),A_{\sigma
}^{(0)}(\sigma ^{\prime },z^{\prime })_{\mathbf{2}}\right] \delta _{\sigma
\sigma ^{\prime }},
\end{aligned}
\end{equation*}
\begin{equation*}
\left\{ A_{\sigma }^{(0)}(\sigma ,z)_{\mathbf{1}},A_{\sigma }^{(2)}(\sigma
^{\prime },z^{\prime })_{\mathbf{2}}\right\} ^{\star} =\left[ R_{%
\mathbf{12}}^{(22)}(z,z^{\prime }),A_{\sigma }^{(2)}(\sigma ,z)_{\mathbf{1}}%
\right] \delta _{\sigma \sigma ^{\prime }}-\left[ R_{\mathbf{12}%
}^{\ast (00)}(z,z^{\prime }),A_{\sigma }^{(2)}(\sigma ^{\prime },z^{\prime
})_{\mathbf{2}}\right] \delta _{\sigma \sigma ^{\prime }},
\end{equation*}
\begin{equation*}
\left\{ A_{\sigma }^{(2)}(\sigma ,z)_{\mathbf{1}},A_{\sigma }^{(0)}(\sigma
^{\prime },z^{\prime })_{\mathbf{2}}\right\} ^{\star} =\left[ R_{%
\mathbf{12}}^{(00)}(z,z^{\prime }),A_{\sigma }^{(2)}(\sigma ,z)_{\mathbf{1}}%
\right] \delta _{\sigma \sigma ^{\prime }}-\left[ R_{\mathbf{12}%
}^{\ast (22)}(z,z^{\prime }),A_{\sigma }^{(2)}(\sigma ^{\prime },z^{\prime
})_{\mathbf{2}}\right] \delta _{\sigma \sigma ^{\prime }}, 
\end{equation*}
\begin{equation}
\begin{aligned}
\left\{ A_{\sigma }^{(2)}(\sigma ,z)_{\mathbf{1}},A_{\sigma }^{(2)}(\sigma
^{\prime },z^{\prime })_{\mathbf{2}}\right\} ^{\star} =-R_{%
\mathbf{12}}^{(22)}(z,&z^{\prime })\delta _{\sigma \sigma ^{\prime }}^{\prime
}+\left[ R_{\mathbf{12}}^{(22)}(z,z^{\prime }),A_{\sigma }^{(0)}(\sigma ,z)_{%
\mathbf{1}}\right] \delta _{\sigma \sigma ^{\prime }} \\
& -R_{\mathbf{12}}^{\ast (22)}(z,z^{\prime })\delta _{\sigma \sigma
^{\prime }}^{\prime }-\left[ R_{\mathbf{12}}^{\ast (22)}(z,z^{\prime
}),A_{\sigma }^{(0)}(\sigma ^{\prime },z^{\prime })_{\mathbf{2}}\right]
\delta _{\sigma \sigma ^{\prime }}.
\end{aligned}
\end{equation}
Finally, the complete covariant Dirac bracket for the gauge fixed theory is
\begin{equation}
\begin{aligned}
\left\{ A_{\sigma }(\sigma ,z)_{\mathbf{1}},A_{\sigma }(\sigma ^{\prime
},z^{\prime })_{\mathbf{2}}\right\} ^{\star} =-R_{\mathbf{12}}(z,z^{\prime })&\delta _{\sigma \sigma^{\prime}}^{\prime }+\left[ R_{%
\mathbf{12}}(z,z^{\prime }),A_{\sigma }(\sigma ,z)_{\mathbf{1}}\right]
\delta _{\sigma \sigma ^{\prime }} \\
&-R_{\mathbf{12}}^{\ast }(z,z^{\prime })\delta _{\sigma \sigma ^{\prime
}}^{\prime }-\left[ R_{\mathbf{12}}^{\ast }(z,z^{\prime }),A_{\sigma
}(\sigma ^{\prime },z^{\prime })_{\mathbf{2}}\right] \delta _{\sigma
\sigma ^{\prime }}. \label{Exchange algebra}
\end{aligned}
\end{equation}
This is the exchange algebra or Maillet bracket \cite{Maillet} of the symmetric space lambda model \cite{k-def}. Notice that there is no $\delta_{zz'}$ or $\hat{\delta}_{zz'}$ term anymore, evidencing that the coordinate $z$ behaves now as an auxiliary variable, i.e. no ultra-locality condition in the $z$ variable is enforced. 

At this  point, $A_{\tau}(z)=\mathscr{L}_{\tau}(z)$ holds strongly but still depending on unknown constants. In order to verify if $A_{\sigma}(z)=\mathscr{L}_{\sigma}(z)$ is valid in the strong sense as well\footnote{We expect this to be the case, as the pair \eqref{pair} is now imposed strongly.}, we evaluate \eqref{Exchange algebra} at $z=z'=z_{\pm}$.  We find two mutually commuting Kac-Moody algebras given by
\begin{equation}
\left\{ A_{\sigma }(\sigma ,z_{\pm})_{\mathbf{1}},A_{\sigma }(\sigma ^{\prime
},z_{\pm})_{\mathbf{2}}\right\} ^{\star} =\mp 2b\left( \left[C_{\mathbf{12}},A_{\sigma}(\sigma',z_{\pm})_{\mathbf{2}} \right]\delta_{\sigma \sigma'}+C_{\mathbf{12}}\delta_{\sigma \sigma'}' \right), \label{Kac-Moody}
\end{equation}
where we have set
\begin{equation}
a=(z_{+}^{4}-z_{-}^{4})/b,
\end{equation}
with $b$ a constant\footnote{The usual WZW model level $k$ can be introduced by taking $b=\pi /k$.}.
If $\mathscr{L}_{\sigma}$ and \eqref{Kac-Moody} are used to find the algebra of $A_{\sigma}(z)$ with itself, we recover \eqref{Exchange algebra} and both are perfectly consistent. This means that the phase space of the reduced theory is described by the data encoded at the points\footnote{This `localization' mechanism in phase space is also observed in the symplectic reduction approach \cite{PCM-Case}.} $z_{\pm}$ via $A_{\sigma}(z_{\pm})$ and that $z$ actually behaves as an spectator parameter, the spectral parameter. The Lax connection can, at this stage, be understood as being valued in the twisted loop algebra defined by%
\begin{equation}
\hat{\mathfrak{f}}=\bigoplus\nolimits_{n\in 
\mathbb{Z}
}\left( \bigoplus\nolimits_{a=0}^{2}\mathfrak{f}^{(a)}\otimes
z^{4n+a}\right) =\bigoplus\nolimits_{n\in 
\mathbb{Z}
}\hat{\mathfrak{f}}^{(n)}.  \label{loop algebra}
\end{equation}
This is the usual starting point used for introducing the Lax pair.

Let us make comment on the computation of \eqref{Kac-Moody}, i.e. the evaluation of \eqref{Exchange algebra} at $z=z'=z_{\pm}$. It is performed by introducing
\begin{equation}
r(z,z^{\prime })_{\mathbf{12}}=\frac{1}{2}\left( R_{\mathbf{12}}(z,z^{\prime })-R_{%
\mathbf{12}}^{\ast }(z,z^{\prime })\right) ,\qquad s(z,z^{\prime })_{%
\mathbf{12}}=\frac{1}{2}\left( R_{\mathbf{12}}(z,z^{\prime })+R_{\mathbf{12}%
}^{\ast }(z,z^{\prime })\right) ,
\end{equation}%
or, equivalently,
\begin{equation}
\begin{aligned}
r(z,z^{\prime })_{\mathbf{12}} &=r_{0}(z,z^{\prime })C_{\mathbf{12}%
}^{(00)}+r_{2}(z,z^{\prime })C_{\mathbf{12}}^{(22)}, \\
s(z,z^{\prime })_{\mathbf{12}} &=s_{0}(z,z^{\prime })C_{\mathbf{12}%
}^{(00)}+s_{2}(z,z^{\prime })C_{\mathbf{12}}^{(22)},
\end{aligned}
\end{equation}%
where%
\begin{equation}
\begin{aligned}
r_{0}(z,z^{\prime }) &=-\frac{1}{z^{4}-z^{\prime 4}}\left( z^{4}\phi
(z)^{-1}+z^{\prime 4}\phi (z^{\prime })^{-1}\right) , \\
r_{2}(z,z^{\prime }) &=-\frac{z^{2}z^{\prime 2}}{z^{4}-z^{\prime 4}}\left(
\phi (z)^{-1}+\phi (z^{\prime })^{-1}\right) , \\
s_{0}(z,z^{\prime }) &=\frac{1}{a}\left( z^{4}+z^{\prime
4}-(z_{+}^{4}+z_{-}^{4})\right) , \\
s_{2}(z,z^{\prime }) &=\frac{1}{a}\frac{1}{z^{2}z^{\prime 2}}\left(
z^{4}z^{\prime ^{4}}-1\right) .
\end{aligned}
\end{equation}%
Form the fact that%
\begin{equation}
r(iz,iz^{\prime })_{\mathbf{12}}=r(z,z^{\prime })_{\mathbf{12}},\text{ \ \ }s(iz,iz^{\prime
})_{\mathbf{12}}=s(z,z^{\prime })_{\mathbf{12}}
\end{equation}%
and
\begin{equation}
s_{\mathbf{12}}(z_{\pm },z_{\pm })=\pm bC_{\mathbf{12}},\qquad s_{%
\mathbf{12}}(z_{\pm },z_{\mp })=r_{\mathbf{12}}(z_{\pm },z_{\mp })=0,\qquad r_{\mathbf{12}}(z_{\pm },z_{\mp })\sim C_{\mathbf{12}},
\end{equation}
we realize that \eqref{Exchange algebra}, actually provides four copies of the Kac-moody algebra \eqref{Kac-Moody}. Denote them by $\left\{ \ast, \ast \right\}_{(\alpha)}$, with \eqref{Kac-Moody} corresponding to $\alpha=0$. This will be useful below.

\subsubsection{Extended lambda model phase space}

Now, we continue with the last step in the Hamiltonian procedure, where the gauge fixing of $P_{\tau}\approx 0$ occurs. We take advantage of this in order to introduce some conditions allowing to determine all the unknown constants in the ansatz \eqref{Ansats for Atau} uniquely, up to a sign.

Besides \eqref{Kac-Moody}, the theory is described by the Poisson brackets corresponding to $\mu=\tau$ in \eqref{canonical PB} and by the total Hamiltonian and constraint given, respectively, by
\begin{equation}
h_{T}=h+\langle u_{\tau},P_{\tau} \rangle_{(\sigma,z)},\qquad P_{\tau}\approx 0.
\end{equation}
For the first class constraint $P_{\tau}$, we choose the following gauge fixing condition
\begin{equation}
A_{\tau}(z)- \mathscr{L}_{\tau}(z)\approx 0. \label{final gauge}
\end{equation} 
We have used the weak symbol $\approx$ to emphasize that the arbitrary constants appearing in the ansatz \eqref{Ansats for Atau} are still to be determined. This is a good gauge fixing condition, whose time preservation determines the Lagrange multiplier $u_{\tau}$. However, as it couples with $P_{\tau}$, its explicit form is of no relevance for the rest of the discussion. The algebra \eqref{Exchange algebra} is not modified by imposing $P_{\tau}=0$ strongly, while the bracket \eqref{canonical PB} for $\mu=\tau$ can be safely discarded.

The reduced theory Hamiltonian is taken to be
\begin{equation}
h=-\frac{ic}{4}
\dint\nolimits_{S^{1}\times \mathbb{CP}^{1}}d\omega \wedge \big\langle A_{\tau }A\big\rangle  \label{red can Ham CS}
\end{equation}
and it is proportional to the `boundary' contribution of the CS canonical Hamiltonian \eqref{can Ham CS}, given by
\begin{equation}
h_{bdry}=\dsum\limits_{\alpha =0}^{3}h_{(\alpha )},\text{ \ \ }h_{(\alpha )}=\frac{1%
}{4b}\dint\nolimits_{S^{1}}d\sigma \left\langle A_{\tau }(i^{\alpha
}z_{+})A_{\sigma }(i^{\alpha }z_{+})-A_{\tau }(i^{\alpha }z_{-})A_{\sigma
}(i^{\alpha }z_{-})\right\rangle, 
\end{equation}
which by virtue of \eqref{comp-trans}, gives
\begin{equation}
h_{bdry}=4h_{(0)},
\end{equation}
with $h_{(0)}=h$ as in \eqref{red can Ham CS}. \\
The Hamiltonian $h_{(0)}$ together with the Kac-Moody algebra $\left\{ \ast, \ast \right\}_{(0)}$, describe correctly the classical dynamics of the symmetric-space lambda model in $F/G$. At this point, we realize that our CS theory is to be defined more appropriately on the orbifold $\Sigma\times \mathbb{CP}^{1}/\mathbb{Z}_{4}$, rather than on the manifold $\Sigma\times \mathbb{CP}^{1}$. This was already noticed in \cite{V3}, from a different perspective. 

When written in the light-cone coordinates\footnote{%
The 2d notation is: $\sigma^{\pm }=\tau\pm \sigma,$ $\partial
_{\pm }=\frac{1}{2}(\partial _{\tau}\pm \partial _{\sigma}),$ $\eta _{\mu \nu
}=diag(1,-1)$ and $a_{\pm }=\frac{1}{2}%
(a_{\tau}\pm a_{\sigma})$.}, \eqref{red can Ham CS} takes the form
\begin{equation}
h=-\frac{ic}{4}
\dint\nolimits_{S^{1}\times \mathbb{CP}^{1}}d\sigma\wedge d\omega \big\langle A_{+}^{2}-A_{-}^{2}\big\rangle.
\end{equation}%
Now, in order to fix the unknown constants in \eqref{Ansats for Atau}, we impose two conditions:\\
i) The quantity
\begin{equation}
p=-\frac{ic}{4}
\dint\nolimits_{S^{1}\times \mathbb{CP}^{1}}d\sigma\wedge d\omega \big\langle A_{+}^{2}+A_{-}^{2}\big\rangle,
\end{equation}%
is conserved and generates $\sigma$ translations along $S^{1}$. This means it is the momentum generator 
\begin{equation}
p=\frac{1}{4b}
\dint\nolimits_{S^{1}} d\sigma \big\langle A_{\sigma}(z_{+})^{2}-A_{\sigma}(z_{-})^{2} \big\rangle.
\end{equation}%
ii) The time evolution dictated by $h$, is such that
\begin{equation}
\left\{h,A_{\sigma}(z) \right\}=\partial_{\sigma}A_{\tau}(z)+\left[A_{\sigma}(z),A_{\tau}(z) \right],\qquad \forall z.
\end{equation}
This is precisely the last eom \eqref{last eom}, but written in Hamiltonian form. \\
Both conditions imply the strongly flatness of the Lax connection from the Kac-Moody algebra structure point of view. They also define the gauge \eqref{final gauge}.  

The condition i) for $A_{\tau}(z_{\pm})$ of the form \eqref{Ansats for Atau}, gives the following relations among the parameters
\begin{equation}
\begin{aligned}
a_{+}(z_{+})a_{-}(z_{+}) &=a_{+}(z_{-})a_{-}(z_{-}),\text{\ \ \ \ \ \ \,}
b_{+}(z_{+})b_{-}(z_{+})=b_{+}(z_{-})b_{-}(z_{-}), \\
a_{+}(z_{+})^{2}-a_{+}(z_{-})^{2} &=1,\text{ \ \ \ \ \ \ \ \ \ \ \ \ \ \ \ \ }b_{+}(z_{+})^{2}-b_{+}(z_{-})^{2}=1, \\
a_{-}(z_{-})^{2}-a_{-}(z_{+})^{2} &=1,\text{ \ \ \ \ \ \ \ \ \ \ \ \ \ \ \
\ }b_{-}(z_{-})^{2}-b_{-}(z_{+})^{2}=1,
\end{aligned}
\end{equation}
while the condition ii), for $z=z_{\pm}$, gives
\begin{equation}
a_{-}(z_{+})=-a_{+}(z_{-}),\qquad b_{-}(z_{+})=-b_{+}(z_{-}).
\end{equation}
The later is precisely the requirement \eqref{vanish bdry}. For generic $z\neq z_{\pm} $, the condition ii) boils down to three conditions according the power of $z$, namely,
\begin{equation*}
\left[ A_{\sigma }^{(0)}(z_{+})-A_{\sigma }^{(0)}(z_{-}),\; A_{\tau
}^{(0)}(z_{+})-A_{\tau }^{(0)}(z_{-})\right] =0,
\end{equation*}%
\begin{equation*}
\begin{aligned}
\Big[ A_{\sigma }^{(0)}(z_{+})-A_{\sigma }^{(0)}(z_{-}),\;& z_{+}^{2}A_{\tau
}^{(2)}(z_{+})-z_{-}^{2}A_{\tau }^{(2)}(z_{-})\Big] \\ 
& +\left[
z_{+}^{2}A_{\sigma }^{(2)}(z_{+})-z_{-}^{2}A_{\sigma }^{(2)}(z_{-}),\; A_{\tau
}^{(0)}(z_{+})-A_{\tau }^{(0)}(z_{-})\right] =0,
\end{aligned}
\end{equation*}
\begin{equation}
\left[ z_{-}^{2}A_{\sigma }^{(2)}(z_{+})-z_{+}^{2}A_{\sigma
}^{(2)}(z_{-}),\; z_{-}^{2}A_{\tau }^{(2)}(z_{+})-z_{+}^{2}A_{\tau
}^{(2)}(z_{-})\right] =0.
\end{equation}
They are enough to fix all the parameters (up to a sign $s=\pm 1$). After some algebra, we obtain
\begin{equation}
\begin{aligned}
a_{+}(z_{+}) &=a_{-}(z_{-})=s,\qquad \qquad \qquad \quad a_{+}(z_{-})=0, \\
b_{+}(z_{+}) &=-b_{-}(z_{-})=s\frac{(z_{+}^{4}+z_{-}^{4})}{%
(z_{+}^{4}-z_{-}^{4})},\qquad b_{+}(z_{-})=s\frac{2}{%
(z_{+}^{4}-z_{-}^{4})},
\end{aligned}
\end{equation}
giving the final form for the time component of the Lax connection,
\begin{equation}
\begin{aligned}
A_{\tau }^{(0)}(z) &=sf_{-}(z)A_{\sigma }^{(0)}(z_{+})+sf_{+}(z)A_{\sigma
}^{(0)}(z_{-}), \\
A_{\tau }^{(2)}(z) &=sg_{-}(z)\frac{z_{+}^{2}}{z^{2}}A_{\sigma
}^{(2)}(z_{+})+sg_{+}(z)\frac{z_{-}^{2}}{z^{2}}A_{\sigma }^{(2)}(z_{-}),
\end{aligned}
\end{equation}
where
\begin{equation}
g_{\pm }(z)=\mp \frac{(z^{4}+z_{\pm }^{4})}{(z_{+}^{4}-z_{-}^{4})}.
\end{equation}

In order to appreciate why $A_{\mu}(z)$ (c.f. \eqref{extended-Lax} above) is the extended symmetric space lambda model Lax pair and not the usual one, we expand the light-cone components $A_{\pm}(z)$ around the zeroes $\mathfrak{z}$ of $\omega$, located at $z=0$ and $z=\infty$. After taking $s=1$, we find
\begin{equation}
A_{+}(z)=I_{\sigma}^{(0)}+z^{2}I_{+}^{(2)}-f_{+}(z)\varphi ^{(0)},\qquad A_{-}(z)=z^{-2}I_{-}^{(2)}, \label{Light-Cone Lax}
\end{equation}%
where
\begin{equation}
I_{\sigma}^{(0)} =A_{\sigma }^{(0)}(z_{+}),\qquad  I_{\pm }^{(2)} =\frac{1}{z_{+}^{4}-z_{-}^{4}}\left( z_{\pm }^{2}A_{\sigma
}^{(2)}(z_{+})-z_{\mp }^{2}A_{\sigma }^{(2)}(z_{-})\right) ,
\end{equation}
are the deformed dual currents and
\begin{equation}
\varphi ^{(0)}=A_{\sigma
}^{(0)}(z_{+})-A_{\sigma }^{(0)}(z_{-}),
\end{equation}
is identified \cite{lambdaCS2} as the first class constraint for the gauge symmetry $G$ of the lambda model on $F/G$. The presence of this term is necessary for the bracket of $A_{\sigma}(z)$ with itself to close into the Maillet algebra form \eqref{Exchange algebra}. Thus, we recover a particular Hamiltonian extension of the ordinary Lax pair. Indeed, in order to achieve the specific form \eqref{extended-Lax} from string theory directly, without having any reference to the CS theory, it is necessary to run the Hamiltonian analysis and extend the usual Lax connection outside the constraint surface by adding to it a combination of the Hamiltonian constraints. See \cite{Magro}, for the case of the $AdS_{5}\times S^{5}$ superstring sigma model. It is remarkable, that it appears in a natural way in the CS theory, as shown above in \eqref{Bdry term}. Furthermore, at this point it is now clear why the ansatz \eqref{Ansats for Atau} was taken. It gives the right form of the Lambda model Lax connection \eqref{Light-Cone Lax}.

Finally, the extended stress tensor density components are given by
\begin{equation}
T_{\pm \pm}=\pm \frac{1}{4b}\big\langle A_{\pm}^{2}(z_{+})-A_{\pm}^{2}(z_{-})  \big\rangle.
\end{equation}
They take the explicit form
\begin{equation}
\begin{aligned}
T_{++}&=\frac{a}{4}\big\langle I_{+}^{(2)} I_{+}^{(2)}\big\rangle-\frac{1}{4b}\big\langle\varphi ^{(0)}\varphi ^{(0)} \big\rangle+\frac{1}{2b}
\big\langle I_{\sigma}^{(0)} \varphi ^{(0)}   \big\rangle, \\
T_{--}&=\frac{a}{4}\big\langle I_{-}^{(2)} I_{-}^{(2)}\big\rangle
\end{aligned}
\end{equation}
and reduce to the usual quadratic coset lambda model stress tensor expressions, when restricted to the surface $\varphi ^{(0)}\approx 0$, where they obey the usual Virasoro algebra.

\section{Concluding remarks}\label{4}

In this note, we have shown how the coset structure of a symmetric space lambda model is implemented in the Chern-Simons approach to integrable systems proposed in \cite{CY}. The reduced theory is described by a particular Hamiltonian extension of the lambda model Lax connection that guarantees the exchange algebra to close in the Maillet algebra form. It is also possible to trace back the origin of the non-ultralocal term $\delta_{\sigma \sigma'}'$ of the lambda model along the gauge fixing procedure. In reverse, that term can be eliminated at the cost of changing the underlying dimensionality of the integrable theory, where the spectral parameter is promoted to a local coordinate of a space that is attached to the 2d world-sheet in the form $\Sigma\times \mathbb{CP}^{1}$ and endowed with the action of a $\mathbb{Z}_{4}$ cyclic group. This raises the possibility of using such a 4d gauge theory reformulation as a tool to tackle the problem of how to quantize non-ultralocal integrable field theories. In particular, integrable (super)-strings in (semi)-symmetric spaces.  

There are some open questions still to be considered. For example: i) how to associate a 2d action functional whose equations of motion are described by the extended Lax connection. This does not seen to be straightforward. As can be observed from the results above, the Lax pair found in \eqref{Light-Cone Lax} is already in a partial gauge constructed directly from the lambda model theory \cite{lambdaCS2}, suggesting that the resulting action functional may include only some of the gauge field components present in the original formulation of the coset lambda models (gauged WZW models, etc). ii) How the lambda model dilaton term contribution to the lambda model action emerge in the process of gauge fixing the CS theory. Perhaps this requires considering manipulations of the CS theory path integral measure. However, as the zeroes of $\omega$ are problematic for defining the theory at the quantum level, how to proceed is still unclear. iii) Recover the $AdS_{5}\times S^{5}$ superstring lambda model from a CS theory defined on a superalgebra. Fortunately, the symmetry structure introduced here extends directly to the supersymmetric case. iv) How to introduce and relate efficiently the action functionals for other integrable deformations of coset string sigma models where Hamiltonian constraints are present (see \cite{Vicedo-Unif} for some examples of construction of explicit actions). Recall that the integrable theory is specified by the poles and zeroes of $\omega$. An analytic continuation relate the poles of $\omega$ for the lambda and eta deformations, which are expected to be related by non-Abelian T-duality \cite{def-drinfeld,E-models}. It would be interesting to address this problem from the CS theory point of view in the coset case. v) How to introduce and arbitrary world-sheet metric. Notice that the results presented here corresponds to the lambda model in conformal gauge. In order to introduce a generic world-sheet metric, the results of \cite{Origin GS} might be useful. \\
Some of these problems are under current investigation.  

\section*{Acknowledgements}

I would like to thank B. Vicedo for comments and suggestions.


\begin{thebibliography}{9}

\bibitem{C1} K. Costello. \textit{Supersymmetric gauge theory and the Yangian}. \href{https://arxiv.org/abs/1303.2632}{[e-Print: arXiv:1303.2632]}. 

\bibitem{C2} K. Costello. \textit{Integrable lattice models from four-dimensional field theories}. Proc.Symp.Pure Math. 88 (2014) 3-24. \href{http://www.ams.org/books/pspum/088/}{https://doi.org/10.1090/pspum/088}. \href{https://arxiv.org/abs/1308.0370}{[e-Print: arXiv:1308.0370]}. 

\bibitem{W} E. Witten. \textit{Integrable lattice models from gauge theory}. Adv. Theor. Math. Phys. 21 (2017) 1819. \href{https://www.intlpress.com/site/pub/pages/journals/items/atmp/content/vols/0021/0007/a010/}{https://dx.doi.org/10.4310/ATMP.2017.v21.n7.a10}. \href{https://arxiv.org/abs/1611.00592}{[e-Print: arXiv:1611.00592]}. 

\bibitem{CWY1} K. Costello, E. Witten and M. Yamazaki. \textit{Gauge theory and integrability, I}. ICCM Not. 6, 46-191 (2018). \href{https://www.intlpress.com/site/pub/pages/journals/items/iccm/content/vols/0006/0001/a006/}{https://dx.doi.org/10.4310/ICCM.2018.v6.n1.a6}. \href{https://arxiv.org/abs/1709.09993}{[e-Print: arXiv:1709.09993]}. 

\bibitem{CWY2} K. Costello, E. Witten and M. Yamazaki. \textit{Gauge theory and integrability, II}.  ICCM Not. 6, 120-149 (2018). \href{https://www.intlpress.com/site/pub/pages/journals/items/iccm/content/vols/0006/0001/a007/}{https://dx.doi.org/10.4310/ICCM.2018.v6.n1.a7}. \href{https://arxiv.org/abs/1802.01579}{[e-Print: arXiv:1802.01579]}. 

\bibitem{CY} K. Costello and M. Yamazaki. \textit{Gauge theory and integrability, III}. \href{https://arxiv.org/abs/1908.02289}{[e-Print: arXiv:1908.02289]}. 

\bibitem{Vicedo-Holo} B. Vicedo. \textit{Holomorphic Chern-Simons theory and affine Gaudin models}. \href{https://arxiv.org/abs/1908.07511}{[e-Print: arXiv:1908.07511]}. 

\bibitem{V3} B. Vicedo. \textit{On integrable field theories as dihedral affine Gaudin models}. Int. Math. Res. Not. \textbf{rny128} (2018). \href{https://academic.oup.com/imrn/article/2020/15/4513/5045617}{https://doi.org/10.1093/imrn/rny128}. \href{https://arxiv.org/abs/1701.04856}{[e-Print: arXiv:1701.04856]}. 

\bibitem{Maillet} J. M. Maillet. \textit{New integrable canonical structures in two-dimensional models}. Nucl. Phys. B269 (1986) 54-76. \href{https://www.sciencedirect.com/science/article/pii/0550321386903652}{https://doi.org/10.1016/0550-3213(86)90365-2}.

\bibitem{Klimcik} Ctirad Klimcik. \textit{Yang-Baxter sigma models and $dS/AdS$ T-duality}. JHEP 0212 (2002) 051. \href{https://iopscience.iop.org/article/10.1088/1126-6708/2002/12/051}{DOI: 10.1088/1126-6708/2002/12/051}. \href{https://arxiv.org/abs/hep-th/0210095}{[e-Print: hep-th/0210095]}. 

\bibitem{eta-def bos} F. Delduc, M. Magro and B. Vicedo. \textit{On classical q-deformations of integrable sigma-models}. JHEP 1311 (2013) 192. \href{https://link.springer.com/article/10.1007/JHEP11(2013)192}{ https://doi.org/10.1007/JHEP11(2013)192}. \href{https://arxiv.org/abs/1308.3581}{[e-Print: arXiv:1308.3581]}.  

\bibitem{eta-def fer} F. Delduc, M. Magro and B. Vicedo. \textit{Integrable deformation of the $AdS_{5} \times S^{5}$ superstring action}. Phys.Rev.Lett. 112 (2014) no.5, 051601. \href{https://journals.aps.org/prl/abstract/10.1103/PhysRevLett.112.051601}{https://doi.org/10.1103/PhysRevLett.112.051601}. \href{https://arxiv.org/abs/1309.5850}{[e-Print: arXiv:1309.5850]}.  

\bibitem{Sfetsos} K. Sfetsos. \textit{Integrable interpolations: from exact CFTs to non-Abelian T-duals}. Nucl.Phys. B880 (2014) 225-246. \href{https://www.sciencedirect.com/science/article/pii/S0550321314000066?via%3Dihub}{https://doi.org/10.1016/j.nuclphysb.2014.01.004}. \href{https://arxiv.org/abs/1312.4560}{[e-Print: arXiv:1312.4560]}. 
 
\bibitem{lambda-bos} T. J. Hollowood, J. L. Miramontes and D. M. Schmidtt. \textit{Integrable deformations of strings on symmetric spaces}. 
JHEP 1411 (2014) 009. \href{https://link.springer.com/article/10.1007%2FJHEP11%282014%29009}{https://doi.org/10.1007/JHEP11(2014)009}. \href{https://arxiv.org/abs/1407.2840}{[e-Print: arXiv:1407.2840]}.  

\bibitem{lambda-fer} T. J. Hollowood, J. L. Miramontes and D. M. Schmidtt. \textit{An integrable deformation of the  $AdS_{5} \times S^{5}$  superstring}.
J.Phys. A47 (2014) no.49, 495402. \href{https://iopscience.iop.org/article/10.1088/1751-8113/47/49/495402}{https://doi.org/10.1088/1751-8113/47/49/495402}. \href{https://arxiv.org/abs/1409.1538}{[e-Print: arXiv:1409.1538]}. 

\bibitem{k-def} T. J. Hollowood, J. L. Miramontes and D. M. Schmidtt. \textit{S-matrices and quantum group symmetry of k-deformed sigma models}. J.Phys.A 49 (2016) 46, 465201. \href{https://iopscience.iop.org/article/10.1088/1751-8113/49/46/465201}{https://doi.org/10.1088/1751-8113/49/46/465201}. \href{https://arxiv.org/abs/1506.06601}{[e-Print: 1506.06601 [hep-th]]}. 

\bibitem{lambdaCS} D. M. Schmidtt. \textit{Integrable lambda models and Chern-Simons theories}. JHEP 1705 (2017) 012. \href{https://link.springer.com/article/10.1007%2FJHEP05%282017%29012}{https://doi.org/10.1007/JHEP05(2017)012}. \href{https://arxiv.org/abs/1701.04138}{[e-Print: arXiv:1701.04138]}. 

\bibitem{lambdaCS2} D. M. Schmidtt. \textit{Lambda models from Chern-Simons theories''}. JHEP 1811 (2018) 111. \href{https://link.springer.com/article/10.1007%2FJHEP11%282018%29111}{https://doi.org/10.1007/JHEP11(2018)111}. \href{https://link.springer.com/article/10.1007%2FJHEP10%282019%29095}{Erratum}. \href{https://arxiv.org/abs/1808.05994}{[e-Print: arXiv:1808.05994]}. 

\bibitem{PCM-Case} D. M. Schmidtt. \textit{Holomorphic Chern-Simons theory and lambda models: PCM case}. JHEP 04 (2020) 060. \href{https://link.springer.com/article/10.1007%2FJHEP04%282020%29060}{https://doi.org/10.1007/JHEP04(2020)060}. \href{https://arxiv.org/abs/1912.07569}{[e-Print: 1912.07569 [hep-th]]}.   

\bibitem{RT} T. Regge, C. Teitelboim. \textit{Role of Surface Integrals in the Hamiltonian Formulation of
General Relativity}. Annals Phys. 88 (1974) 286. \href{https://www.sciencedirect.com/science/article/pii/0003491674904047}{https://doi.org/10.1016/0003-4916(74)90404-7}.

\bibitem{BH1} J. D. Brown, M. Henneaux. \textit{Central Charges in the Canonical Realization of Asymptotic
Symmetries: An Example from Three-Dimensional Gravity}. Commun. Math. Phys. 104 (1986). \href{https://link.springer.com/article/10.1007%2FBF01211590}{https://doi.org/10.1007/BF01211590}.
207.

\bibitem{BH2} J. D. Brown, M. Henneaux. \textit{On the Poisson Brackets of Differentiable Generators in
Classical Field Theory}. J. Math. Phys. 27 (1986) 489. \href{https://aip.scitation.org/doi/10.1063/1.527249}{https://doi.org/10.1063/1.527249
}.

\bibitem{B1} M. Ba\~nados. \textit{Global charges in Chern-Simons field theory and the $(2+1)$ black hole}. Phys. Rev.
D 52 (1996) 5816. \href{https://journals.aps.org/prd/abstract/10.1103/PhysRevD.52.5816}{https://doi.org/10.1103/PhysRevD.52.5816}. \href{https://arxiv.org/abs/hep-th/9405171}{[e-Print: arXiv:9405171 [hep-th]]}.

\bibitem{B2} M. Ba\~nados. \textit{Three-dimensional quantum geometry and black holes}. AIP Conf. Proc. 484, no.1
(1999) 147. \href{https://aip.scitation.org/doi/abs/10.1063/1.59661}{https://doi.org/10.1063/1.59661
}. \href{https://arxiv.org/abs/hep-th/9901148}{[e-Print: arXiv:9901148
 [hep-th]]}.

\bibitem{BR} M. Ba\~nados, I. A. Reyes. \textit{A short review on Noethers theorems, gauge symmetries and
boundary terms}. Int. J. Mod. Phys. D 25 (2016) no.10, 1630021. \href{https://www.worldscientific.com/doi/abs/10.1142/S0218271816300214}{https://doi.org/10.1142/S0218271816300214}. \href{https://arxiv.org/abs/1601.03616}{[e-Print: arXiv:1601.03616 [hep-th]]}.

\bibitem{Magro} M. Magro. \textit{The classical exchange algebra of $AdS_{5}\times S^{5}$}. JHEP 01 (2009) 021. \href{https://iopscience.iop.org/article/10.1088/1126-6708/2009/01/021}{https://doi.org/10.1088/1126-6708/2009/01/021}. \href{https://arxiv.org/abs/0810.4136}{[e-Print: 0810.4136 [hep-th]]}. 

\bibitem{Vicedo-Unif} F. Delduc, S. Lacroix, M. Magro and B. Vicedo. \textit{A unifying 2d action for integrable $\sigma$-models from 4d Chern-Simons theory}. \href{https://link.springer.com/article/10.1007%2Fs11005-020-01268-y}{https://doi.org/10.1007/s11005-020-01268-y}. \href{https://arxiv.org/abs/1909.13824}{[e-Print: arXiv:1909.13824]}. 

\bibitem{def-drinfeld} B. Vicedo \textit{Deformed integrable $\sigma$-models, classical R-matrices and classical exchange algebra on Drinfeld doubles}. J.Phys. A48 (2015) no.35, 355203. \href{https://iopscience.iop.org/article/10.1088/1751-8113/48/35/355203}{https://doi.org/10.1088/1751-8113/48/35/355203}. \href{https://arxiv.org/abs/1504.06303}{[e-Print: arXiv:1504.06303]}. 

\bibitem{E-models} Ctirad Klimcik. \textit{$\eta$ and $\lambda$ deformations as $E$-models}. Nucl.Phys. B900 (2015) 259-272. \href{https://www.sciencedirect.com/science/article/pii/S0550321315003302?via%3Dihub}{https://doi.org/10.1016/j.nuclphysb.2015.09.011}. \href{https://arxiv.org/abs/1508.05832}{[e-Print: arXiv:1508.05832]}. 

\bibitem{Origin GS} K. Costello, B. Stefa\'nski. \textit{Chern-Simons origin of superstring integrability}. Phys.Rev.Lett. 125 (2020) 12, 121602. \href{https://journals.aps.org/prl/abstract/10.1103/PhysRevLett.125.121602}{https://doi.org/10.1103/PhysRevLett.125.121602}. \href{https://arxiv.org/abs/2005.03064}{[e-Print: 2005.03064 [hep-th]]}. 

\end{thebibliography}
\end{document}